\documentclass[aps,pre,preprint,amsmath,amssymb,nofootinbib]{revtex4-2}

\usepackage[T1]{fontenc}
\usepackage{graphicx}
\usepackage{hyperref}
\usepackage{xcolor}
\definecolor{darkred}{RGB}{139,0,0}
\hypersetup{colorlinks=true, linkcolor=darkred, citecolor=darkred, urlcolor=darkred}

\makeatletter\@booleanfalse\preprintsty@sw\makeatother

\newcommand{\ind}[1]{\mathbf{1}_{#1}}

\newcommand{\der}{\text{d}}

\makeatletter
\@tfor\sok@l:=abcdefghijklmnopqrstuvwxyz\do{%
  \expandafter\edef\csname bb\sok@l\endcsname{\noexpand\boldsymbol{\sok@l}}%
}
\@tfor\sok@l:=ABCDEFGHIJKLMNOPQRSTUVWXYZ\do{%
  \expandafter\edef\csname bb\sok@l\endcsname{\noexpand\boldsymbol{\sok@l}}%
}
\makeatother

\makeatletter
\@tfor\sok@l:=ABCDEFGHIJKLMNOPQRSTUVWXYZ\do{%
  \expandafter\edef\csname \sok@l\sok@l\endcsname{\noexpand\mathbb{\sok@l}}%
}
\makeatother

\newcommand{\lr}[1]{\left(#1\right)}
\newcommand{\lrb}[1]{\left[#1\right]}
\newcommand{\lrc}[1]{\left\{#1\right\}}

\usepackage{amsthm}

\theoremstyle{definition}

\usepackage{amsmath,amssymb,amsthm}
\usepackage[margin=2.5cm]{geometry}

\theoremstyle{remark}

\makeatletter
\let\jstat@makecaption\@makecaption
\long\def\@makecaption#1#2{{\def\baselinestretch{0.9}\jstat@makecaption{#1}{#2}}}
\makeatother

\begin{document}

\title{Pushy Random Walk on the Bethe lattice:\\ a Galton--Watson perspective}
\author{J\'er\'emie Klinger}
\affiliation{Laboratoire de Physique de l'\'Ecole normale sup\'erieure, ENS, Universit\'e PSL}

\begin{abstract}
    Pushy random walks are minimal models of transport in deformable media, in which the walker displaces the obstacles it bumps into rather than being blocked by them.
    On the Bethe lattice, we show that the walkable structure dynamically created by a pushy walker can be generated a priori as an explicit multi-type Galton--Watson tree, so that the walker reduces to a simple random walk on a static random tree, and its escape to infinity coincides with the survival of the underlying branching process.
    Standard generating-function tools then give direct access to the escape properties of these walks.
    For the Sokoban walk, which pushes a single obstacle, we recover the known critical density $\rho_c = 1-(z-1)^{-2}$ and the behavior of the escape probability close to $\rho_c$, and derive its small-density asymptotics.
    For the second-order pushy walk, which can push trains of two obstacles, we establish the conjectured $\rho_c = 1-(z-1)^{-3}$, and compute the escape probability at all densities.
    Beyond escape, the same framework yields the statistics of the trace of trapped walkers: the distribution of the maximal depth follows from the extinction probabilities, while the distributions of the visited volume and of the number of dead ends are encoded in explicit functional equations for their generating functions, which we invert numerically.
    All predictions agree with Monte Carlo simulations.
\end{abstract}

\maketitle

\section{Introduction}

Random walks on graphs \cite{hughes1995random1,weiss1994aspects,lyons2016probability} provide a natural framework for transport in disordered media.
In their simplest form, disorder is introduced by randomly placing obstacles that the walker cannot traverse, i.e.\ by site or bond percolation of an underlying graph \cite{stauffer1994introduction}, which yields the lattice Lorentz gas, or ``ant in a labyrinth'' \cite{degennes1976percolation,lorentz1905mouvement}.
Occupying each node independently with probability $\rho$ by an inert, immovable obstacle intuitively slows down -- and can even trap -- the walker, and precisely characterizing this effect has been the subject of intense research on regular, fractal and hyperbolic geometries alike \cite{gefen1983anomalous,halperin1985differences,kesten1986subdiffusive,grimmett1993random,barlow2004random,benavraham2000diffusion,lyons2016probability}.
Two questions are then central. The first is whether the walker escapes to infinity with positive probability, $\PP\lrb{\bbx_n\to \infty}>0$, where $\bbx_n$ denotes its graph distance from the origin after $n$ steps.
The second concerns the \textit{trace} (or \textit{range} \cite{dvoretzky1951some,jain1972range}) of the walk, that is, the properties of the set of sites it eventually visits, viewed as a random geometrical object.
For instance, provided the walker does not escape, how far does it travel, and how many sites does it visit?
For the Lorentz gas, both questions are purely \textit{static}: the walker eventually visits every site of the cluster of free sites containing the origin, so that escape amounts to the percolation \cite{stauffer1994introduction} of this cluster, and the statistics of the trace are those of the cluster.

Lorentz gases have proven undeniably successful in modeling diffusive to subdiffusive empirical transport \cite{bruin1972logarithmic,james1987diffusion,hofling2008critical,krakoviack2009tagged,murase2004ultrafine,nicolau2007sources,hofling2013anomalous}. However, the underlying assumption of immovable obstacles is open to debate. In active microrheology for instance, a probe, typically driven through biological material, might rearrange the surrounding crowded medium \cite{meyer2006laser,biswas2020first,demery2014generalized}. These effects are not confined to microscales: invading tumour cells remodel the surrounding tissue \cite{friedl2003tumour}, and robots exploring arenas of movable obstacles showcase similar environment rearrangements \cite{altshuler2024environmental}.
To better model such empirical relizations, Bonomo and Reuveni introduced the pushy -- or \textit{Sokoban} -- random walk as an insightful alternative paradigm, in which the walker can displace obstacles \cite{bonomo2023loss}.
A persistent version of the Sokoban walk was subsequently shown to capture some of the essential physics of the experiments \cite{altshuler2024environmental}, and the model has since been investigated in a variety of settings \cite{bonomo2024bethe,singh2026crossover,singh2026trapping,bonomo2026pushy,shitrit2026arrest}.
Such walks belong to the broader family of random walks that modify their environment as they move, such as excited random walks \cite{benjamini2003excited}.
Specifically, when attempting to hop onto a node occupied by an obstacle, the Sokoban walker pushes the obstacle to the next node, provided that node is itself free of obstacles.
In contrast to Lorentz gases, in which the environment is drawn at random once and for all at initialization, the Sokoban walk dynamically modifies this random initial condition, generating long-lived correlations between the walker and the graph geometry.
In particular, the trace is no longer a pre-existing cluster of the environment, but is built by the walker itself.

The resulting dynamics were first studied on $\ZZ^2$, where, numerical simulations surprisingly indicate that the percolation property is lost; for any initial obstacle density $\rho>0$, the walker dynamically builds a cage that traps it, so that the escape probability vanishes almost surely \cite{bonomo2023loss,singh2026crossover}.
However, the looped structure of the two-dimensional lattice makes the problem particularly hard, and has so far prevented an exact proof.
More recently, the Sokoban walk has been studied on the Bethe lattice \cite{bonomo2024bethe} -- the infinite regular tree in which every node has degree $z$ (see Fig.~\ref{fig:bethe_mgw}).
This choice is natural and important: since the Bethe lattice has no loops, many problems that are intractable in finite dimension become exactly solvable, and mean-field approximations become exact.
As such, it has long served as the reference mean-field geometry in statistical physics, from percolation \cite{fisher1961cluster,coniglio1976some} to spin glasses \cite{thouless1986spin,mezard2001bethe} and Anderson localization \cite{abouchacra1973selfconsistent}.
In this setting, it was proven \cite{bonomo2024bethe} that the escape probability is strictly positive for obstacle densities below an explicit critical threshold $\rho < \rho_c = 1 - (z-1)^{-2}$.
In words, Sokoban walkers can push their way out of a congested Bethe lattice, provided the congestion level remains below $\rho_c$.
Remarkably, this threshold is found both from a mean-field argument and an exact calculation.
It has further been conjectured that walkers able to push trains of $k$ obstacles, hereafter denoted PRW$_k$ (Pushy Random Walk of order $k$, so that the Sokoban walk is PRW$_1$, while the RW in a Lorentz gas, which cannot push obstacles, is PRW$_0$), escape with positive probability below $\rho_c = 1 - (z-1)^{-(1+k)}$.
However, the calculations employed in the introductory work \cite{bonomo2024bethe} are specifically tailored to the Sokoban case, and do not transfer to this extended setting.
Furthermore, the geometry of the region explored by the Sokoban walker has, to our knowledge, not been investigated.

In this article, we develop a framework that overcomes these limitations.
Our key observation is that, on the Bethe lattice, the dynamical graph explored by a pushy walker can be mapped onto a random tree generated by an explicit multi-type Galton--Watson (MGW) process \cite{watson1875probability,harris1963theory,athreya1972branching}.
This completely decouples the random dynamics of the walker from the random environment: the pushy walk becomes a nearest-neighbor walk on a static random tree.
Beyond its conceptual simplicity, this mapping gives direct access to the extensive mathematical literature on MGW processes \cite{harris1963theory,athreya1972branching,lyons2016probability}: the escape probability is related to extinction probability of the tree, and the trace becomes a finite MGW tree, whose statistics are amenable to analysis by generating function techniques.
Using this mapping, we show that $\rho_c = 1-(z-1)^{-3}$ for PRW$_2$, thereby settling the conjecture of \cite{bonomo2024bethe} for $k=2$.
We further compute escape probabilities for all $\rho<\rho_c$, and obtain the full distributions of the maximal depth, the volume and the number of dead ends of the trace for $\rho\geq\rho_c$. 
In principle, the method extends to larger values of $k$, but involves combinatorics of increasing complexity, which we leave for future work.

We begin in Section~\ref{sec:lorentz} with the Lorentz gas on the Bethe lattice. While this section contains no new results, the mapping onto a single-type Galton--Watson (GW) process is straightforward in this case, and allows us to build intuition and to introduce standard GW tools and terminology.
In Section~\ref{sec:prw1}, we turn to the Sokoban walk itself: we show how the walkable structure is revealed as the walker explores the lattice, construct the corresponding MGW process, and use it to rederive and extend all the results of \cite{bonomo2024bethe}.
Section~\ref{sec:prw2} applies the same methodology to PRW$_2$, and establishes its critical density and escape probabilities.
In Section~\ref{sec:traces}, we leverage the MGW mapping beyond escape probabilities, and investigate some statistics of the trace of PRW$_1$ and PRW$_2$ for $\rho\geq\rho_c$.
In particular, we consider the distribution of the maximal depth, the total number of visited sites and the number of dead ends, all of which are implicitly given by explicit functional equations satisfied by their generating functions.
Throughout, MGW predictions are compared with direct Monte Carlo simulations of PRW$_1$ and PRW$_2$.

\section{Lorentz gas on the Bethe lattice}
\label{sec:lorentz}

\subsection{Random walk on the Bethe lattice}

\begin{figure}[h!]
    \includegraphics[height=0.25\textwidth]{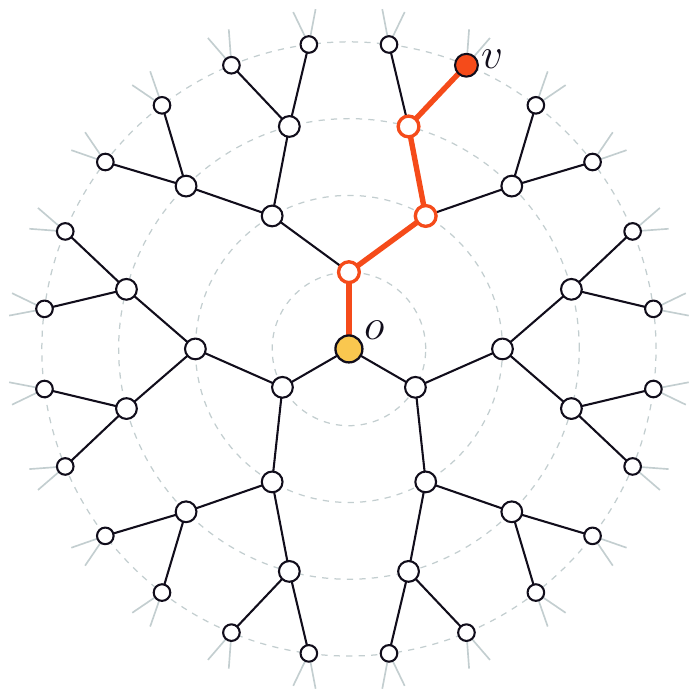}
    \includegraphics[height=0.25\textwidth]{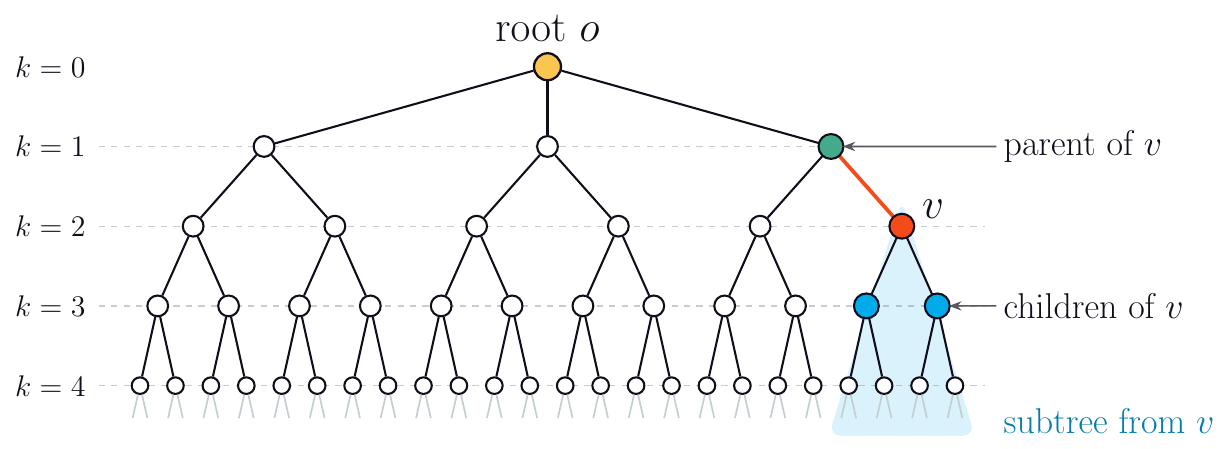}
    \caption{{\bf (left)} Cropped Bethe lattice of coordination $z=3$. The isotropic nature of the graph suggests this natural radial representation, where the walker starts from the root $o$. For a given node $v$ (in red), there is a single path linking $o$ and $v$.~{\bf (right)} Alternative genealogical tree representation: the root $o$ is now the oldest ancestor and level $k$ gathers the nodes at distance $k$ from the root. The genealogical terminology is easily grasped within this vertical picture.}
    \label{fig:bethe_mgw}
\end{figure}

In graph terminology, the Bethe lattice \cite{bethe1935statistical} of coordination $z$ (see Fig.~\ref{fig:bethe_mgw} (left) for the case $z=3$) is a hyperbolic tree.
Every node $u$ of the graph has exactly $z$ neighbors, the number $N_d(u)$ of nodes at distance $d$ from $u$ grows exponentially fast, as $N_d(u)\propto (z-1)^{d}$, and the structure is loopless, that is, for any other node $v$ there is a single path linking $u$ and $v$.
Beyond the natural radial representation of the Bethe lattice,  it is often convenient to designate a given node as the \textit{root} $o$ of the lattice, with the rest of the structure hanging from it, as depicted in Fig.~\ref{fig:bethe_mgw} (right). We then pictorially refer to the directions towards and away from the root as upward and downward, respectively.
\textcolor{black}{This vertical picture naturally comes with the genealogical terminology of Galton--Watson trees \cite{watson1875probability}, which we use throughout. The $z-1$ downward neighbors of a node $v$ are its \textit{children}, and its unique upward neighbor is its \textit{parent}.
Note that the root is the only node with no parent and $z$ children.
The \textit{descendants} of $v$ are all the nodes that can be reached from $v$ by moving downward only, and, together with $v$, they form the \textit{subtree hanging from} $v$. Nodes at distance $k$ from the root form the $k$th \textit{generation} of the lattice, which contains $z(z-1)^{k-1}$ nodes for $k\geq 1$. Finally, in any subtree of the lattice, a node with no children in that subtree is called a \textit{leaf}.}

Consider a random walk (RW) on the Bethe lattice, starting from the root $o$, and denote by $\bbx_n$ its distance from the root after $n$ steps, so that $\bbx_0 = 0$.
At each integer time step $n$, the walker moves to one of its $z$ neighbors, chosen uniformly at random with probability $z^{-1}$, and one may ask about its long-time behavior.
For any $z\geq 3$, it turns out that the RW on the Bethe lattice is \textit{transient} \cite{lyons2016probability}: the expected number of returns to the root is finite and, consequently, the probability of escaping to infinity, $\PP\lrb{\bbx_n\to \infty}$, is one \cite{lyons2016probability}. Moreover, the escape is actually ballistic, that is,

\begin{equation}
    \underset{n\to \infty}{\lim}\frac{\bbx_n}{n} = \frac{z-2}{z}.
\end{equation}
This is to be contrasted with the more standard diffusive behavior $\bbx_n\propto \sqrt{n}$ of symmetric nearest-neighbor random walks on a line.
In words, the imbalance between the $z-1$ neighbors leading away from the root and the unique neighbor leading back induces an entropic downward drift, which drives the ballistic escape to infinity.

\subsection{Effective graph of the Lorentz gas}

\begin{figure}[h!]
    \includegraphics[width=0.8\textwidth]{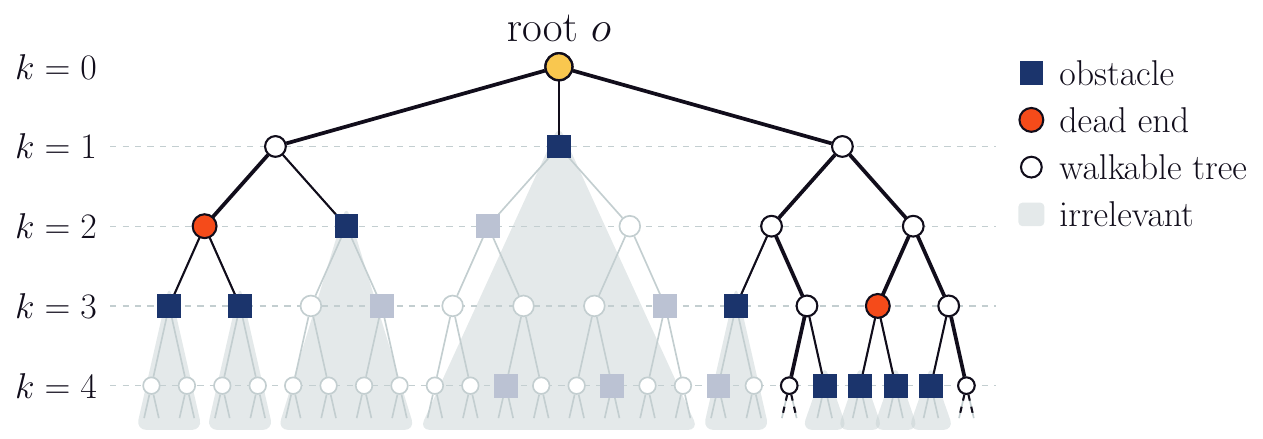}
    \caption{A potentially infinite Lorentz gas configuration on the Bethe lattice. Starting from the root $o$, a RW can only visit non-obstructed nodes, and is completely blind to the obstacle configurations in the subtrees of obstructed nodes. 
    From the GW perspective, it is clear that the root has a slightly modified offspring distribution; in this realization, it has two children.}
    \label{fig:lorentz_bethe}
\end{figure}

We now build the Lorentz gas on the Bethe lattice. On each node of the graph (except the root), we place an inert obstacle with probability $\rho \in [0,1]$ (see Fig.~\ref{fig:lorentz_bethe}). The resulting graph is random, and the trajectory of the walker, hereafter denoted PRW$_0$, is constrained by the forbidden nodes.
Because of the tree structure of the Bethe lattice, the obstacle configuration below any obstructed node $v$ is completely irrelevant to the RW, since it cannot bypass an obstacle at $v$ in any way.
Moreover, if all children of an accessible node $v'$ carry an obstacle, a RW reaching $v'$ can only backtrack from this effective dead end.
From the point of view of the RW, the walkable graph is thus a random tree with leaves (dead ends), and the statistics of this tree govern the behavior of the RW.

In fact, this effective tree is most easily constructed using the general theory of Galton--Watson (GW) processes \cite{watson1875probability,harris1963theory,athreya1972branching}.
Let $Z_k$ denote the random number of nodes accessible to the RW at distance $k$ from the root, and suppose the RW arrives at one of them and attempts to travel further downward. From the Lorentz gas construction, it is clear that the number $Z$ of accessible \textit{children} follows a binomial distribution, $Z \sim \mathcal{B}(z-1, 1-\rho)$.
As a result, the number of reachable nodes at distance $k+1$ is given by the recursive relation

\begin{equation}
    \label{eq:gw_lorentz}
    Z_{k+1} =\sum_{i=1}^{Z_k} Z^{(i)}, ~~~~ Z^{(i)} \overset{\text{i.i.d.}}{\sim} Z \sim \mathcal{B}(z-1, 1-\rho),
\end{equation}
which defines a Galton--Watson process for $k\geq 1$. The distribution of $Z_1 \sim \mathcal{B}(z, 1-\rho)$ is slightly different, since by construction of the Bethe lattice the root can have up to $z$ children.
Following GW terminology, the accessible children of a node are its \textit{offspring}, the $k$th generation of the GW process is made of the accessible nodes at distance $k$ from the root, and the realization of the process is a random \textit{GW tree}, corresponding to a random subtree of the Bethe lattice hanging from the root. Its leaves are precisely the dead ends of the Lorentz gas, and the GW tree is finite, or \textit{extinct}, if and only if some generation is empty, $Z_k = 0$.

\

\subsection{Random walk on Galton--Watson trees}

The GW tree \eqref{eq:gw_lorentz} is an alternative description of the random structure seen by a RW traveling through the Lorentz gas on the Bethe lattice. Importantly, the statistics of $Z_k$ dictate the long-time behavior of the RW.
Consider a given GW tree and assume, for instance, that $Z_k = 0$ for some $k$.
This event clearly has nonzero probability, since $\PP\lrb{Z_1 = 0} = \rho^{z} > 0 $.
Let now a RW evolve on this finite tree; the RW obviously remains bounded and can never escape to infinity.
Conversely, defining the $\rho$-dependent GW extinction probability $q^o = \PP\lrb{\exists k\geq 1, Z_k = 0}$, the probability that the RW escapes to infinity is given by \cite{lyons2016probability} 

\begin{equation}
    \label{eq:esc_lorentz}
    \PP^{\text{PRW}_0}\lrb{\bbx_n \to \infty} = 1 - q^o.
\end{equation}
Note that on the l.h.s.\ the probability is taken over realizations of PRW$_0$ in the Lorentz gas on the Bethe lattice, while the r.h.s.\ is a pure GW quantity, typically referred to as the survival probability.

A large body of work deals with the extinction and survival of GW processes \cite{harris1963theory,athreya1972branching,lyons2016probability}; we here recall the tools needed later for the analysis of pushy walkers.
Assume for a moment that the root has the same offspring distribution as any member of the GW process, that is, $Z_0 = 1, Z_1 \sim Z$, and denote by $q$ the corresponding GW extinction probability.
Of utmost importance in GW theory is the \textit{probability generating function (pgf)} $f$, which equivalently defines the distribution of $Z$:

\begin{equation}
    \label{eq:pgf_gw}
    f(s) = \EE\lrb{s^Z} = \sum_k \PP\lrb{Z=k}s^k.
\end{equation}
For the Lorentz gas, it is clear from Fig.~\ref{fig:lorentz_bethe} that $Z \sim \mathcal{B}(z-1, 1-\rho)$, so that $f$ takes the explicit form
\begin{equation}
    \label{eq:pgf_lorentz}
    f(s) = \lr{\rho + (1-\rho)s}^{z-1}.
\end{equation}
By virtue of normalization, $f(1)=1$. 
Moreover, the mean offspring number is given by $m = f'(1) = (z-1)(1-\rho)$, and yields the expected number of descendants at the $k$th generation,

\begin{equation}
    \label{eq:mean_gw}
    \EE\lrb{Z_{k}} = m^k.
\end{equation}
The mean offspring number $m$ separates almost-sure extinction from survival with positive probability. For $m<1$, $\EE\lrb{Z_{k}}$ decays exponentially fast and a GW tree is almost surely finite \cite{athreya1972branching}, meaning that $q = 1$. For $m>1$, $\EE\lrb{Z_{k}}$ grows exponentially fast. Note that this does not guarantee survival (recall that $\PP\lrb{Z_1 = 0} = \rho^{z-1} > 0$). However, the extinction probability is then strictly less than one. 
More precisely, $q < 1$ is given by the smallest solution in $[0,1]$ of the fixed-point equation \cite{athreya1972branching}

\begin{equation}
    \label{eq:fixed_point_gw}
    q = f(q).
\end{equation}
With the explicit pgf \eqref{eq:pgf_lorentz}, Eq.~\eqref{eq:fixed_point_gw} is a polynomial equation of degree $z-1$ in $q$, which in general has to be solved numerically by fixed-point iteration.
As a side note, the specific case $z=3$ is exactly solvable, with extinction probability given by $q = \min\lr{1, \rho^2/(1-\rho)^2}$ (see Appendix~\ref{app:z3}).
Finally, the critical threshold corresponds to $m=1$, which in terms of obstacle density simply reads
\begin{equation}
    \label{eq:crit_gw}
    m = 1 \Leftrightarrow \rho_c = 1 - \frac{1}{z-1}.
\end{equation}

\begin{figure}[h!]
    \includegraphics[width=0.5\textwidth]{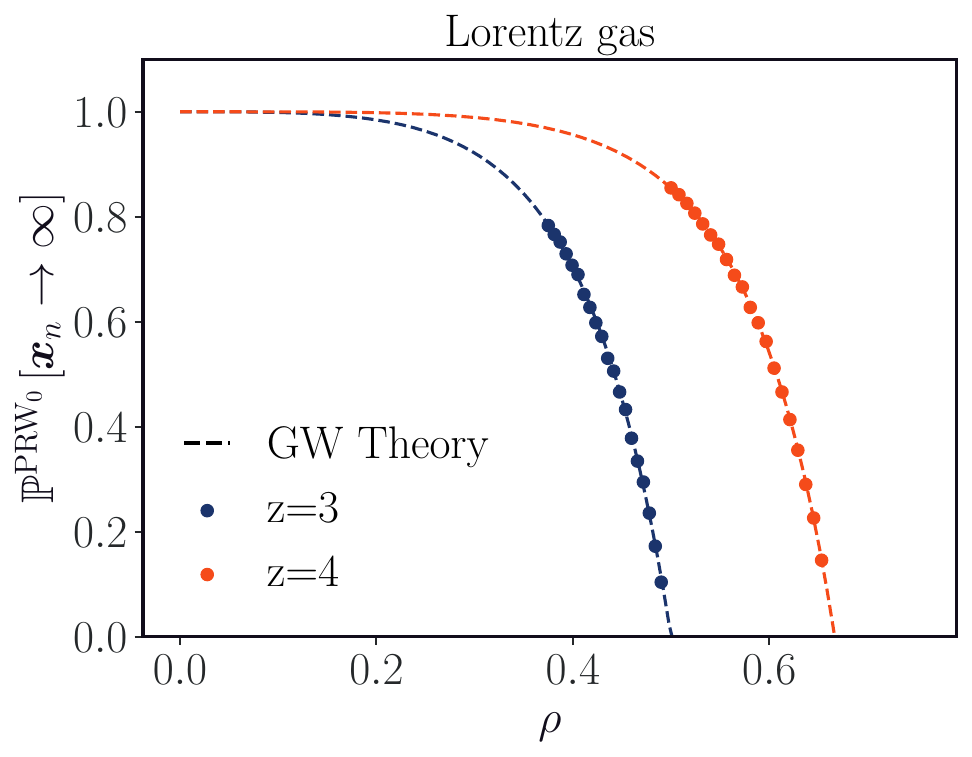}
    \caption{
    Escape probabilities of PRW$_0$ in the Lorentz gas on the Bethe lattice, at varying coordination number $z$. 
    Monte Carlo simulations are averaged over $2\times 10^4$ trials, where each RW is evolved for $10^6$ steps and counts as escaped if the maximum root distance does not saturate, i.e.\ keeps increasing over the last $5\times 10^5$ steps.
    The theoretical dashed lines are obtained from the numerical procedure discussed following Eq.~\eqref{eq:esc_lorentz_final}.
    The critical threshold is apparent for both $z=3,4$ values.
    }
    \label{fig:p0_esc}
\end{figure}

Returning to the RW navigating the Lorentz gas on the Bethe lattice, the escape probability is obtained from minor modifications of these generic GW results. 
While the critical threshold is unchanged, the escape probability \eqref{eq:esc_lorentz} must take into account the modified offspring distribution of the root. Defining the true pgf of the root, $f^o(s) = \EE\lrb{s^{Z_1}} = \lr{\rho + (1-\rho)s}^{z}$, the escape probability is given in terms of $q$: 

\begin{equation}
    \label{eq:esc_lorentz_final}
    \PP^{\text{PRW}_0}\lrb{\bbx_n \to \infty} = 1 - q^o = 1 - f^o(q).
\end{equation}
In the (sub)critical case $\rho \geq \rho_c$, extinction occurs almost surely, so that $q=1$ and $\PP^{\text{PRW}_0}\lrb{\bbx_n \to \infty}=0$. 
In the supercritical case $\rho < \rho_c$, the escape probability is obtained by combining the fixed-point solution of Eq.~\eqref{eq:fixed_point_gw} with the explicit root pgf $f^o$.
As shown in Fig.~\ref{fig:p0_esc}, these numerical solutions agree seamlessly with direct simulations of PRW$_0$ on the Bethe lattice.
For $z=3$, the escape probability is fully explicit and given in Appendix~\ref{app:z3}.

As announced in the introduction, the escape probabilities in Fig.~\ref{fig:p0_esc} are not new \cite{bonomo2024bethe}, but the GW route provides an alternative and straightforward derivation.
At first glance, determining the escape probabilities of pushy random walks seems to require different tools. 
We show in the next section that this is not the case, provided one maps the problem onto the survival of multi-type GW processes.

\section{Sokoban walk on the Bethe lattice}
\label{sec:prw1}
\label{sec:perco}

\begin{figure}[h!]
    \includegraphics[width=\textwidth]{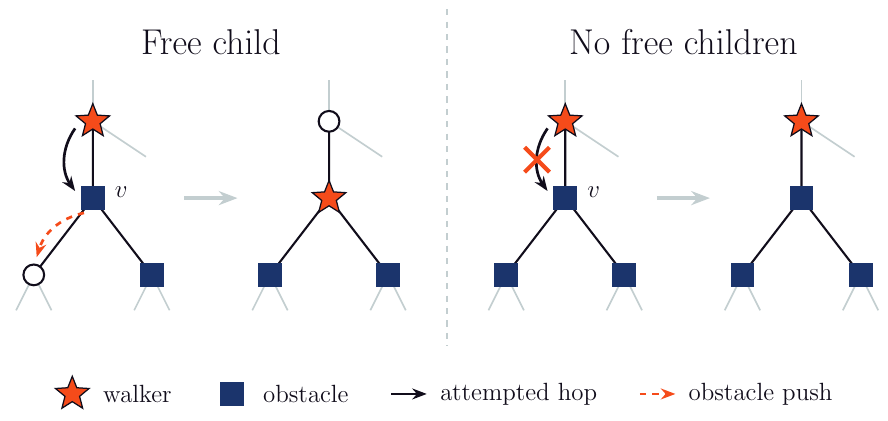}
    \caption{Displacement rules of the Sokoban walk PRW$_1$ ($z=3$). For each case, the walker attempts to hop onto an obstructed node $v$. {\bf (left)} At least one child of $v$ is free: the obstacle is pushed onto a free child, chosen uniformly at random, and the walker moves to $v$. {\bf (right)} All the children of $v$ are obstructed: the hop fails and the walker stays put.}
    \label{fig:rule_prw1}
\end{figure}

Departing from PRW$_0$ in a Lorentz gas on the Bethe lattice, we now consider the Sokoban walk, PRW$_1$, with the following modified dynamics.
Starting from the same Lorentz initial condition, with a density $\rho$ of obstacles on the lattice, the walker performs nearest-neighbor hopping, and pushing occurs whenever it attempts to hop onto an obstructed node $v$.
If at least one child of $v$ is free, the obstacle on $v$ is pushed onto one of the free children, chosen uniformly at random, and the walker moves to $v$.
Otherwise, all the children of $v$ are obstructed, the hop fails and the walker stays put (see Fig.~\ref{fig:rule_prw1} for a schematic of the situation).

\subsection{Revealing the walkable tree}

Thanks to the loopless structure of the Bethe lattice, obstacles can only be displaced when PRW$_k$ moves away from the root. Thus, obstacles only travel downward, and the walker never encounters obstacles when backtracking along its trajectory.
As a consequence, when the walker reaches a node $v$ for the first time, the subtree hanging from $v$ has never been visited, and its configuration only differs from the initial Lorentz one through the obstacles pushed into it from above.
In other words, whenever the walker attempts to visit a node for the first time, this node reveals its accessible descendants, according to a law that only depends on the state of the node when it is reached, which we call its \textit{type}, and on fresh, independent randomness.

Since each revelation only involves independent randomness, the order in which nodes are revealed is irrelevant: rather than following the walker, one can equivalently generate the walkable structure once and for all from the Lorentz gas initialization, level by level from the root.
This is precisely the growth rule of a multi-type GW process, with one type per kind of node.
The PRW$_k$ then becomes a simple RW on a static random tree with leaves, as for the Lorentz gas in Section~\ref{sec:lorentz}. Of course, the random motion of the walker implies that it may take a fair amount of time to uncover the entire structure; however, the statistical properties of this structure do not depend on how it is uncovered and, moreover, they govern the escape probability of PRW$_k$.
In the remainder of this section, we explicitly discuss node types and construct the equivalent MGW process for PRW$_1$.

\subsection{Node types and MGW process for PRW$_1$}

The analysis of the walkable structure generated by PRW$_1$ requires two distinct types.
Let $v$ be a freshly discovered node. If $v$ is free, which we call type $s$, then each of its children carries an obstacle with probability $\rho$.
However, if $v$ is occupied, which we call type $m$, then either at least one of its children is free and the obstacle on $v$ is displaced there, or all the children of $v$ are occupied, in which case $v$ is a dead end: PRW$_1$ can never travel through $v$ and must backtrack.

Following the level-by-level construction above, the iterative growth of the walkable structure is formalized by a multi-type GW (MGW) process.
As for the Lorentz gas, the $k$th generation of the MGW process consists of nodes at distance $k$ from the root, and each node now carries a type, which determines the law of its typed offspring. Contrary to the Lorentz gas, however, the MGW tree also contains nodes that PRW$_1$ can never enter, namely $m$-type nodes whose children are all of type $m$. 
While this subtlety is irrelevant for computing escape probabilities, it plays a role in trace statistics, as further discussed in Section~\ref{sec:traces}.

As in Section~\ref{sec:lorentz}, we first ignore the special role of the root, and denote by MGW$_i$, $i\in\{s,m\}$, the MGW process started from a single node of type $i$.
Let us define the random vector $\bbZ_k = (Z_k^s, Z_k^m)^T$, which counts the nodes of each type at generation $k$, and let $|\bbZ_k| = \bbZ_k^T\ind{}$ be the total number of offspring at generation $k$. With $\bbZ_0 = (1,0)^T$ for MGW$_s$ and $\bbZ_0 = (0,1)^T$ for MGW$_m$ respectively, for any $k\geq 0$, $\bbZ_k$ satisfies the random recursion

\begin{equation}
    \label{eq:recur_prw1}
        \bbZ_{k+1} = \sum_{i=1}^{Z_k^s}\bbX^{s,(i)} + \sum_{j=1}^{Z_k^m}\bbX^{m,(j)}~~\text{with}~~\bbX^{s,(i)}, \bbX^{m,(j)}\overset{\text{i.i.d}}{\sim}\bbX^s,\bbX^m,
\end{equation}
where the random typed offspring vectors are explicitly given by

\begin{equation}
    \label{eq:offspring_prw1}
    \bbX^s = \begin{pmatrix}
        z-1-\xi\\ \xi
    \end{pmatrix}~~~~~ \bbX^m = \ind{\xi < z - 1}\begin{pmatrix}
        z-2-\xi\\ 1 + \xi
    \end{pmatrix}~~~~\xi \sim \mathcal{B}\lr{z-1, \rho}.
\end{equation}
The indicator function $\ind{\xi < z-1}$ specifically accounts for the fact that $m$-type nodes have no viable offspring if all their children are themselves of type $m$. Otherwise, the obstacle on $v$ is pushed onto one of its free children, which thereby becomes of type $m$.
Here and throughout, superscripts refer to node types: $\bbX^i$ is the offspring vector of a type-$i$ node, whose component $X^i_j$ counts its type-$j$ children, and pgfs $f^i$, extinction probabilities $q^i$, probabilities $\PP^i$ and expectations $\EE^i$ similarly refer to a process started from a single type-$i$ node.

\subsection{Extinction probabilities for MGW trees}

As in the single-type case, probability generating functions are the tool of choice to determine extinction probabilities, and we recall here standard results on MGW processes \cite{harris1963theory,athreya1972branching}.
For any $\bbs = (s_1, s_2) \in \RR^2$, the $s$- and $m$-type probability generating functions are defined as

\begin{equation}
    \label{eq:pgf_prw1}
    f^s(\bbs) = \EE\lrb{s_1^{X^s_s} s_2^{X^s_m}}, ~~ f^m(\bbs) = \EE\lrb{s_1^{X^m_s} s_2^{X^m_m}},
\end{equation}
and we introduce the joint probability generating function $\bbf(\bbs) = (f^s(\bbs), f^m(\bbs))^T$.
As in the single-type case, $\bbf$ fully characterizes the MGW process. In particular, the mean offspring matrix

\begin{equation}
    M = \text{Jac}^T(\bbf)|_{\bbs = \ind{}};~~~~M_{ij} = \EE\lrb{X^j_i}
\end{equation}
yields the mean offspring counts at generation $k$,
\begin{equation}
    \EE\lrb{\bbZ_k|\bbZ_0} = M^k \bbZ_0,
\end{equation}
and its largest Perron root $\varrho(M)$ controls extinction \cite{harris1963theory,athreya1972branching}.
Denote by $\bbq = (q^s, q^m)^T$ the extinction probability vector, where $q^i = \PP^i\lrb{\exists k \geq 0; |\bbZ_k| = 0}$ is the probability that the lineage of a single type-$i$ node dies out.
For $\varrho(M) \leq 1$, extinction occurs almost surely, $\bbq = \ind{}$.
When $\varrho(M) > 1$, the process survives with positive probability, $q^s, q^m < 1$, and $\bbq$ is the smallest solution in $[0,1]^2$ of the fixed-point equation

\begin{equation}
    \label{eq:fp_prw1}
    \bbq = \bbf(\bbq).
\end{equation}

\subsection{Escape probability of PRW$_1$ on the Bethe lattice}

We now return to the Bethe lattice, on which the root plays a special role: it is of type $s$, but has $z$ children instead of $z-1$ (recall Fig.~\ref{fig:lorentz_bethe}) so that $\bbZ_1 \sim (z- \xi, \xi)^T$ with $\xi \sim \mathcal{B}\lr{z, \rho}$.
The walkable structure of PRW$_1$ is therefore the slightly refined process MGW$_o$, whose first generation is drawn from this modified distribution, and which then follows the recursion \eqref{eq:recur_prw1} for $k\geq 1$.
As in the Lorentz gas, this refinement does not require any extra computation. 
Denoting by $f^o(\bbs) = \EE\lrb{s_1^{Z_1^s} s_2^{Z_1^m}}$ the probability generating function of the root, the extinction probability of MGW$_o$ is directly obtained from $\bbq$:

\begin{equation}
    q^o = \PP^o\lrb{\exists k \geq 0; |\bbZ_k| = 0} = f^o(\bbq),
\end{equation}
and the escape probability of PRW$_1$ corresponds to the survival probability of MGW$_o$

\begin{equation}
    \label{eq:esc_prw1}
    \PP^{\text{PRW}_1}\lrb{\bbx_n \to \infty} = 1 - q^o = 1 - f^o(\bbq).
\end{equation}
In particular, $\PP^{\text{PRW}_1}\lrb{\bbx_n \to \infty}=0$ whenever $\varrho(M) \leq 1$, $\PP^{\text{PRW}_1}\lrb{\bbx_n \to \infty}>0$ whenever $\varrho(M) > 1$, and criticality corresponds to $\varrho(M) = 1$.

\begin{figure}[h!]
    \includegraphics[width=0.5\textwidth]{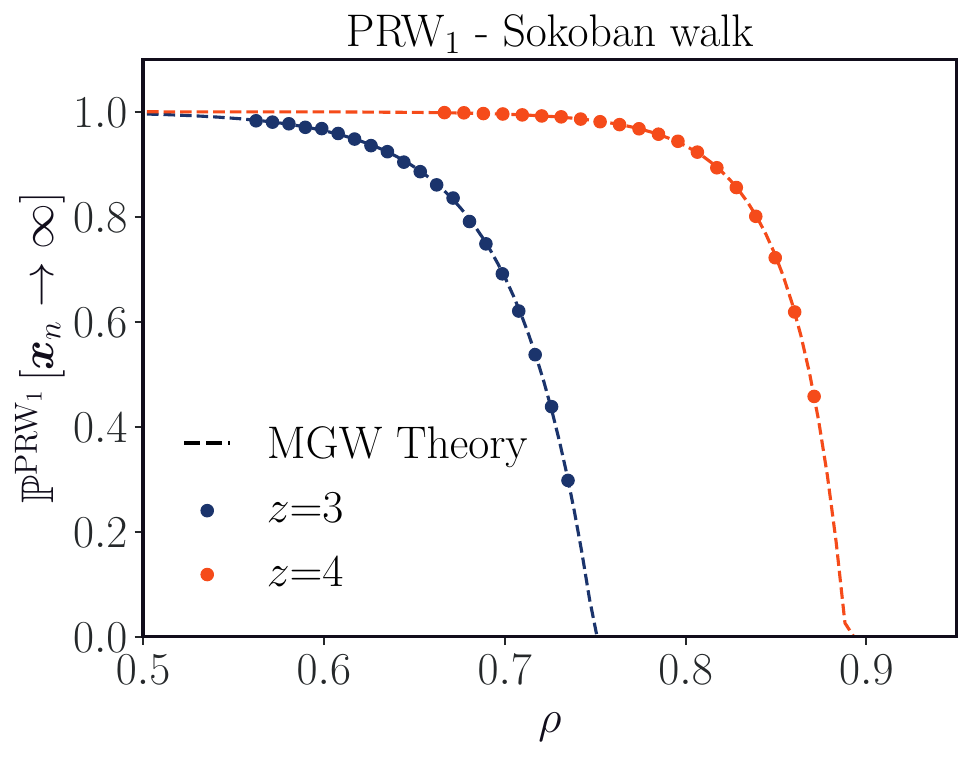}
    \caption{
    Escape probabilities of PRW$_1$ on the Bethe lattice, at varying coordination number $z$. 
    Monte Carlo simulations are averaged over $2\times 10^4$ trials, where each RW is evolved for $10^6$ steps and counts as escaped if the maximum root distance does not saturate, i.e.\ keeps increasing over the last $5\times 10^5$ steps.
    The theoretical dashed lines are obtained from the numerical procedure discussed following Eq.~\eqref{eq:crit_prw1}.
    The critical threshold is apparent for both $z=3,4$ values.
    }
    \label{fig:p1_esc}
\end{figure}

The explicit typed offspring distributions \eqref{eq:offspring_prw1} allow us to specialize the generic MGW results discussed above to PRW$_1$. A direct computation shows that the $(s,m,o)$ pgfs are given by

\begin{equation}
    \label{eq:exp_pgf_prw1}
    \begin{split}
        &f^s(\bbs) = \lrb{(1-\rho)s_1 + \rho s_2}^{z-1}\\
        &f^m(\bbs) = s_2 s_1^{z-2}\lrb{1-\rho + \frac{\rho s_2}{s_1}}^{z-1} + \rho^{z-1}\lrb{1 - \frac{s_2^z}{s_1}}\\
        &f^o(\bbs) = \lrb{(1-\rho)s_1 + \rho s_2}^{z}\\
    \end{split}
\end{equation}
and the corresponding mean offspring matrix reads

\begin{equation}
    \label{eq:mom_prw1}
    M(\rho,z) = \begin{pmatrix}
        (z-1)(1-\rho) & (z-1)(1-\rho) + \rho^{z-1} - 1\\
        (z-1)\rho & (z-1)(\rho-\rho^{z-1}) + 1-\rho^{z-1}\\
    \end{pmatrix}.
\end{equation}
Together, the pgfs \eqref{eq:exp_pgf_prw1} and mean offspring matrix \eqref{eq:mom_prw1} allow us to fully analyze the escape probability of PRW$_1$.
In particular, the critical density is obtained by finding $\rho$ such that the largest Perron root of $M$ is equal to 1. The characteristic polynomial $P_{\rm car}(\lambda)\equiv\text{Det}\lrb{M - \lambda \II_2}$ at $\lambda=1$ is explicit and given by

\begin{equation}
    \label{eq:cp_prw1}
    P_{\rm car}(1) = (z-1)^2\rho^{z-1}\lrb{\rho - \lr{1 - \frac{1}{(z-1)^2}}},
\end{equation}
from which one immediately recovers the critical density \cite{bonomo2024bethe}
\begin{equation}
    \label{eq:crit_prw1}
    \rho_c = 1 - \frac{1}{(z-1)^2} 
\end{equation}
that is, the only nonzero density at which $1$ is a Perron root of $M(\rho, z)$. As a result, for (sub)critical $\rho \geq \rho_c$, $\bbq = \ind{}$ and
PRW$_1$ escapes to infinity with zero probability, $\PP^{\text{PRW}_1}\lrb{\bbx_n\to \infty} = 0$.
For supercritical $\rho  < \rho_c$, the two-dimensional fixed-point equation \eqref{eq:fp_prw1} must in general be solved numerically, using the explicit pgfs \eqref{eq:exp_pgf_prw1},
and Fig.~\ref{fig:p1_esc} displays seamless agreement between fixed-point solutions and Monte Carlo simulations of PRW$_1$.
As a side note, the case $z=3$ is again exactly solvable, and the corresponding $\bbq$ is explicitly given in Appendix~\ref{app:z3}.

\subsection{Perturbative $\rho$ analysis}

\begin{figure}[h!]
    \includegraphics[width=0.48\textwidth]{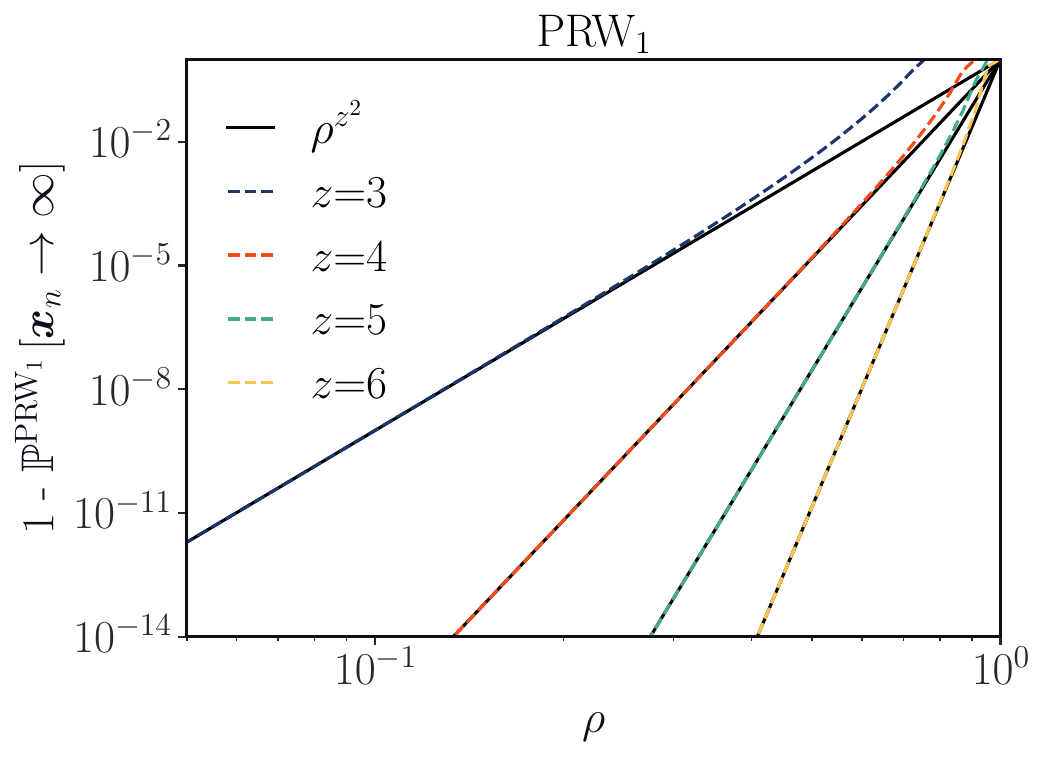}
    \includegraphics[width=0.48\textwidth]{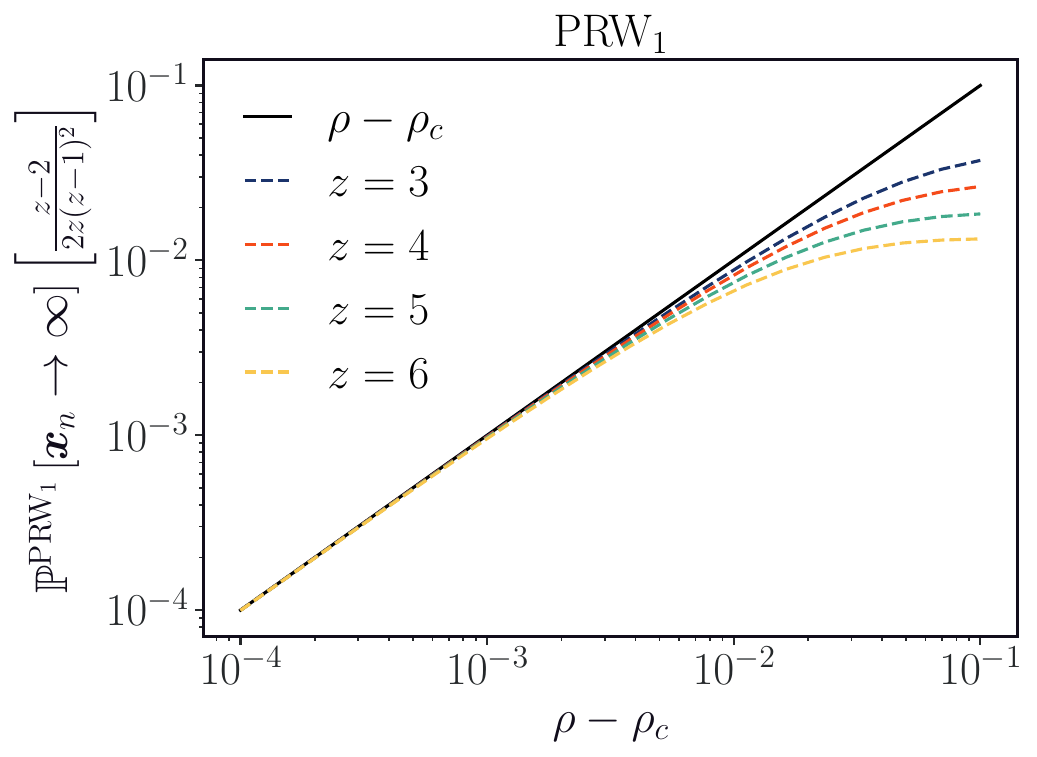}
    \caption{Perturbative behavior of the escape probability of PRW$_1$, across coordination number $z$. In both panels, solid black lines are the first-order predictions and coincide with the dashed lines obtained by numerically solving Eqs.~\eqref{eq:fp_prw1}--\eqref{eq:esc_prw1}. The curves deviate away from $\rho = 0$ and $\rho = \rho_c$ respectively. \textbf{(left)} Small-$\rho$ behavior of the trapping probability $1-\PP^{\text{PRW}_1}\lrb{\bbx_n\to\infty}$. \textbf{(right)} Close to $\rho_c$, the escape probability is linear in $(\rho_c-\rho)$ with explicit $z$-dependent slope \eqref{eq:derivative_prw1}; all curves collapse once rescaled.}
    \label{fig:p1_rho_pert}
\end{figure}

The explicit pgfs \eqref{eq:exp_pgf_prw1} allow for a perturbative analysis of escape probabilities close to both $\rho=0$ and $\rho = \rho_c$. 
The vicinity of $\rho = 0$ is the easiest: starting from \eqref{eq:exp_pgf_prw1}, we expect $\bbq \sim \boldsymbol{0}$ and plug in the extinction probability ansatz $\bbq = (\alpha_c \rho^\alpha, \beta_c \rho^\beta)$, where $(\alpha_c, \alpha, \beta_c, \beta) \in \RR_+^4$ are positive constants to be determined:

\begin{equation}
    \begin{split}
        &\alpha_c \rho^\alpha = \lrb{(1-\rho) \alpha_c \rho^\alpha + \beta_c \rho^{\beta + 1}}^{z-1}\\
        &\beta_c \rho^\beta =  \beta_c \alpha_c^{z-2} \rho^{\beta + \alpha (z-2)}\lrb{1-\rho + \frac{  \beta_c \rho^{\beta + 1 - \alpha}}{\alpha_c}}^{z-1} + \rho^{z-1}\lrb{1 - \frac{ \beta_c^z \rho^{\beta z - \alpha}}{\alpha_c }}.
    \end{split}
\end{equation}
To lowest order in $\rho$, the second equality reads
\begin{equation}
    \beta_c \rho^\beta = \beta_c \alpha_c^{z-2} \rho^{\beta + \alpha (z - 2)} + \rho^{z-1},
\end{equation}
and, recalling that $z\geq 3$, we obtain $(\beta_c, \beta) = (1, z-1)$. 
In turn, the first equality yields
\begin{equation}
    \alpha_c \rho^\alpha = \lrb{(1-\rho) \alpha_c \rho^\alpha + \rho^{z}}^{z-1},
\end{equation}
which, to lowest order in $\rho$, gives $(\alpha_c, \alpha) = (1, z(z-1))$. Making use of the explicit root pgf $f^o$, the small-$\rho$ behavior of the escape probability is then given by

\begin{equation}
    \label{eq:smallrho_prw1}
    1 - \PP^{\text{PRW}_1}\lrb{\bbx_n \to \infty} \underset{\rho \to 0}{\sim} \rho^{z^2},
\end{equation}
as numerically illustrated in Fig.~\ref{fig:p1_rho_pert} (left).

The analysis around $\rho_c$ is a paradigmatic example of a perturbative expansion carried out close to a singularity, and requires more care.
To avoid ambiguity, we explicitly write out the $\rho$ dependence of both $\bbq$ and $\bbf$, and aim to compute the $\rho$ derivative at criticality, $\partial_\rho \bbq(\rho)|_{\rho_c} \equiv \bbp$.
Considering the small parameter $\epsilon = \rho_c - \rho$, and evaluating the fixed-point equation \eqref{eq:fp_prw1} at $\rho = \rho_c - \epsilon$ yields

\begin{equation}
    \bbq(\rho_c - \epsilon) = \bbf(\bbq(\rho_c - \epsilon),  \rho_c - \epsilon).
\end{equation}
Recalling that $\bbq(\rho_c) = \ind{}$, denoting the Jacobian by $J(\bbq, \rho) = \nabla_{\bbs} \bbf(\bbs, \rho)|_{\bbs = \bbq}$, and expanding to first and second order in $\epsilon$, we obtain the system

\begin{equation}
    \label{eq:expansion_prw1}
    \begin{split}
        &O(\epsilon):~~~~\lrb{J(\ind{}, \rho_c) - \II_2}\bbp = 0\\
        &O(\epsilon^2):~~~~\frac{1}{2}\lrb{J(\ind{}, \rho_c) - \II_2}\partial_{\rho} \bbp = \partial_\rho J(\ind{}, \rho_c)\bbp + \frac{1}{2}B[\bbp]
    \end{split}
\end{equation}
where we use that $\partial^{(k)}_\rho \bbf(\ind{}, \rho) = 0$ since for any density $\bbf(\ind{}, \rho) = \ind{}$, and we denote by $B[\bbs]$ the quadratic form $B[\bbs]_i = \sum_{kl}s_k s_l \partial_{x_k, x_l}f^i|_{\bbx = \ind{}}$.

The $O(\epsilon)$ equation does not determine $\bbp$ uniquely, but constrains it to be proportional to a right null eigenvector of $J-\II_2$, which exists by definition of criticality. Such a vector is actually explicit, and we obtain that $\bbp = \alpha (z, 1)^T \equiv \alpha \bbu$ with $\alpha$ some unknown constant.
The second equation determines $\alpha$: letting $\bbv$ be an explicit left null eigenvector of $J-\II_2$ and taking the scalar product of the $O(\epsilon^2)$ equation with $\bbv$ yields
\begin{equation}
    0 = \alpha \lrb{2\bbv^T \partial_\rho J(\ind{}, \rho_c)\bbu + \alpha \bbv^T B\lrb{\bbu}} \implies \alpha = - \frac{2\bbv^T \partial_\rho J(\ind{}, \rho_c)\bbu}{\bbv^T B[\bbu]} =  \frac{2(z-1)^3}{(z-2)z}.
\end{equation}
In turn, the derivative of the escape probability of PRW$_1$ at $\rho = \rho_c$ is finally obtained as
\begin{equation}
    \label{eq:derivative_prw1}
    \partial_\rho \PP^{\text{PRW}_1}\lrb{\bbx_n\to\infty}|_{\rho_c} = - \lrb{\nabla_{\bbs} f^o(\ind{}, \rho_c)}^T\bbp = - \frac{2 z (z-1)^2}{z-2},
\end{equation} 
and the limiting linear behavior of $\PP^{\text{PRW}_1}\lrb{\bbx_n\to\infty}$ close to $\rho_c$ is numerically illustrated in Fig.~\ref{fig:p1_rho_pert} (right).

In conclusion, the MGW mapping successfully recovers the critical density \eqref{eq:crit_prw1} and the behavior at small $\rho_c - \rho$ \eqref{eq:derivative_prw1} already derived in \cite{bonomo2024bethe}, while providing further insight into the small-$\rho$ behavior \eqref{eq:smallrho_prw1} of the escape probability of PRW$_1$, as well as straightforward fixed-point schemes to evaluate it at any $\rho < \rho_c$.
Most importantly, this mapping and the associated techniques are designed to generalize, and we show in the next section that PRW$_2$ can be investigated similarly.

\section{Second-order pushy random walk on the Bethe lattice}
\label{sec:prw2}

Since the revelation argument of Section~\ref{sec:prw1} holds for any $k$, and the MGW toolkit has been discussed at length there, we directly construct the MGW process equivalent to the second-order pushy random walk, PRW$_2$. 
We first recall the obstacle displacement rules in this case, which are also illustrated in Fig.~\ref{fig:rules_prw2} of Appendix~\ref{app:rules_prw2}.
Whenever PRW$_2$ attempts to hop onto an occupied node $v$, the corresponding obstacle is displaced downward. It either occupies a free child of $v$, or pushes the obstacle of an occupied child one level further down, provided there is a free node there. 
In both cases, each displaced obstacle moves to a child chosen uniformly among the admissible ones, that is, children that are either free, or obstructed but with at least one free child of their own.
The hop fails whenever all the descendants of $v$ in the two layers below it are obstructed, which occurs with probability $\rho^{z(z-1)}$.

\subsection{Node types and MGW process for PRW$_2$}

\begin{figure}[h!]
    \includegraphics[width=\textwidth]{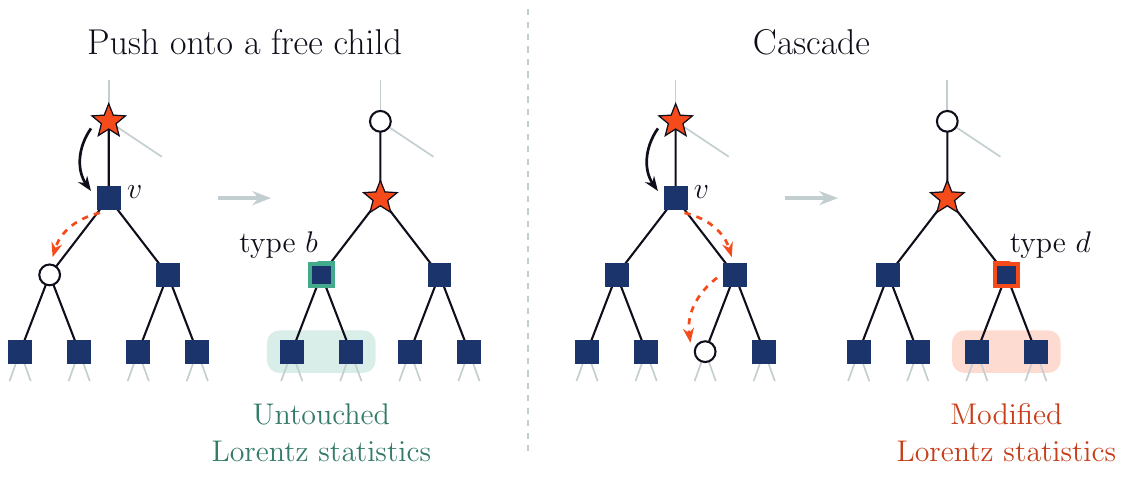}
    \caption{Pushing mechanism and corresponding node types generated by PRW$_2$. {\bf (left)} Upon hopping onto $v$, the walker displaces the obstacle onto a free child. The grandchildren of $v$ are unaffected by this event, and the node that received the obstacle is either of type $f$ or $b$ (here, $b$). {\bf (right)} In this scenario, the obstacle of $v$ cascades to one of its grandchildren. The child involved in the cascade is thus of type $d$, as the obstacle distribution of its own children has been modified by the cascading event.}
    \label{fig:prw2_types}
\end{figure}

Consider now the various states a node $v$ can be in when PRW$_2$ first attempts to hop onto it.
What matters for the subsequent evolution is whether $v$ is occupied, and whether the occupation of its children still follows the initial Lorentz statistics.
The easiest case, which we call \textit{type} $s$, is that of a free node: its children have never been affected by a pushing event, and each of them independently carries an obstacle with probability $\rho$.
Assume now that $v$ is occupied. Whether its obstacle is the one from the initial Lorentz configuration, or has been pushed onto $v$ by a previous displacement, is irrelevant: what matters is whether $v$'s own children have been affected by such a displacement, as illustrated in Fig.~\ref{fig:prw2_types}.
This only happens when $v$ sits in the middle of a cascade, that is, when the obstacle initially on $v$ has been pushed onto one of its free children, and replaced by the obstacle coming from its parent.
If the children of $v$ are untouched, it is useful to distinguish two types: we call \textit{type} $f$ an occupied node with at least one free child, and \textit{type} $b$ an occupied node whose children are all obstructed.
Otherwise, we call \textit{type} $d$ an occupied node that sits in the middle of a previous cascade.
The $f/b$ distinction is useful since an occupied node with untouched children is of type $f$ or $b$ with probability $1-\rho^{z-1}$ or $\rho^{z-1}$, respectively.
Furthermore, such conditional information will prove necessary to decide whether an MGW node of a given type may have no children. 

We now translate these scenarios into explicit typed random offspring distributions, with the convention $\bbX^i = (X^i_s,X^i_f,X^i_b,X^i_d)^T$ for the offspring vector of a type-$i$ node.
\begin{enumerate}
    
    \item Type $s$ is the simplest, since entering a free node induces no downward obstacle cascade:
\begin{equation}
    \begin{split}
        \label{eq:rvs_prw2}
    &\bbX^s =(\xi^s,  \xi^f,  \xi^b, 0 )^T\\
     &\xi\equiv (\xi^s, \xi^f, \xi^b)^T \sim \text{Multinomial}\lrb{z-1, \lr{1-\rho, \rho(1 - \rho^{z-1}), \rho \rho^{z-1}}}.
    \end{split}
\end{equation}
    \item Type $b$ is also straightforward: all its children are occupied and untouched, hence of type $f$ or $b$. If they are all of type $b$, the node is a dead end. Otherwise, entering the node necessarily triggers a cascade through one of its $f$-type children, which thereby becomes of type $d$:
\begin{equation}
    \label{eq:rvb_prw2}
    \begin{split}
    &\bbX^b = \ind{\xi < z-1 }(0,  z- 2 - \xi,  \xi,  1)^T
    ~~~\xi \sim \mathcal{B}\lrb{z-1, \rho^{z-1}}    .
    \end{split}
\end{equation}

    \item Type $f$ is trickier: the node cannot be a dead end, since we know it has at least one free child. However, its obstacle can be pushed either onto a free child, which becomes an occupied node with untouched children, that is, of type $f$ or $b$, or onto an $f$-type child, which becomes of type $d$ as its own obstacle is pushed one level further down. The corresponding offspring distribution reads
    
    \begin{equation}
        \label{eq:rvf_prw2}
        \begin{split}
            &\bbX^f = \begin{pmatrix}
        \xi^s \\ \xi^f \\ \xi^b \\ 0
    \end{pmatrix} + u \begin{pmatrix}
        - 1 \\ 1 - \chi  \\ \chi \\ 0
    \end{pmatrix} + (1-u)\begin{pmatrix}
        0 \\ -1 \\ 0 \\ 1
    \end{pmatrix}\\
&\xi\equiv (\xi^s, \xi^f, \xi^b)^T \sim \text{Multinomial}\lrb{z-1, \lr{1-\rho, \rho(1 - \rho^{z-1}), \rho \rho^{z-1}}| \xi^s > 0}\\
&u\sim\text{Ber}\lrb{\frac{\xi^s}{\xi^s + \xi^f}}, \chi \sim \text{Ber}\lrb{\rho^{z-1}} 
        \end{split}
\end{equation}
where the random variable $u$ accounts for whether a node of type $s$ or $f$ receives the displaced obstacle. 

\item Finally, type $d$ is the most complicated. Since it sits in the middle of a previous cascade, we already know that it initially had at least one free child, and that one of its free children has since received an obstacle, thereby becoming of type $f$ or $b$. Accounting for this prior information, as well as for the additional obstacle displaced when PRW$_2$ hops onto the node, yields the random offspring vector

\begin{equation}
    \label{eq:rvd_prw2}
        \begin{split}
            &\bbX^d = \ind{\xi^b + \chi < z-1}\lrb{\begin{pmatrix}
        \xi^s - 1 \\ \xi^f + 1-\chi \\ \xi^b + \chi\\ 0
    \end{pmatrix} + u \begin{pmatrix}
        - 1 \\ 1 - \chi_2  \\ \chi_2 \\ 0
    \end{pmatrix} + (1-u)\begin{pmatrix}
        0 \\ -1 \\ 0 \\ 1
    \end{pmatrix}}\\
&\xi\equiv (\xi^s, \xi^f, \xi^b)^T \sim \text{Multinomial}\lrb{z-1, \lr{1-\rho, \rho(1 - \rho^{z-1}), \rho \rho^{z-1}}| \xi^s > 0}\\
&\chi, \chi_2 \sim \text{Ber}\lrb{\rho^{z-1}}, u\sim\text{Ber}\lrb{\frac{\xi^s - 1}{\xi^s - 1 + \xi^f + 1 - \chi}}
        \end{split}
\end{equation}
\end{enumerate}
Together, \eqref{eq:rvs_prw2}--\eqref{eq:rvd_prw2} define the MGW process underlying the walkable structure dynamically created by PRW$_2$.

\subsection{Critical density}

In spirit, the determination of the critical density of PRW$_2$ follows that of PRW$_1$. 
While the mean offspring matrix can be computed explicitly (see Appendix~\ref{app:mom_prw2}), its characteristic polynomial does not admit a factorization as simple as Eq.~\eqref{eq:cp_prw1}.
We therefore detail an alternative representation of the mean offspring matrix, from which $\rho_c$ follows explicitly.
To lighten expressions, we introduce the notations $(n,a,b,c,r,S) = (z-1,\ 1-\rho,\ \rho(1-\rho^{z-1}),\ \rho \rho^{z-1},\ \rho^{z-1},\ 1-\rho^{z-1})$, and investigate each mean offspring vector separately.

Again, types $s$ and $b$ are the simplest, and the multinomial and binomial laws imply
\begin{equation}
    \label{eq:mov1_prw2}
    \EE\lrb{\bbX^s} = \begin{pmatrix}
    na \\ nb \\ nc \\ 0
    \end{pmatrix}, ~~~~\EE\lrb{\bbX^b} = \begin{pmatrix}
        0\\ n(1 - r) - (1 - r^n) \\ n (r - r ^n) \\1 - r^n
    \end{pmatrix}.
\end{equation}
To tackle types $f$ and $d$, it is convenient to split each offspring vector into a part that does not depend on $u$, and a part associated with the obstacle pushing event encoded by the random variable $u$. The $f$-type mean offspring vector then reads

\begin{equation}
    \label{eq:movf_prw2}
    \EE\lrb{\bbX^f} = \begin{pmatrix}
        \frac{na}{S} \\ \frac{nb(1-\rho^{n-1})}{S} - 1 \\ \frac{nc(1-\rho^{n-1})}{S} \\ 1
    \end{pmatrix} + \pi^f \begin{pmatrix}
        -1 \\ 2-r \\ r \\ -1
    \end{pmatrix} \equiv \overline{\EE\lrb{\bbX^f}} + \pi^f \bbd
\end{equation}
where $\bbd$ is a fixed vector and $\pi^f = \EE\lrb{u}$, with $u$ defined in the $f$-type random offspring vector \eqref{eq:rvf_prw2}.
The $d$ type is dealt with similarly,

\begin{equation}
    \label{eq:movd_prw2}
    \EE\lrb{\bbX^d} =  \begin{pmatrix}
        \frac{na}{S} -1 \\ \frac{nb(1-\rho^{n-1})}{S} - r + p \\ \frac{nc(1-\rho^{n-1})}{S} +r - pn\\ 1-p
    \end{pmatrix} +  \pi^d \bbd \equiv  \overline{\EE\lrb{\bbX^d} }  + \pi^d \bbd
\end{equation}
where, under the laws \eqref{eq:rvd_prw2}, $p =\PP\lrb{\xi^b = n-1, \chi=1|\xi^s>0} = \frac{n c^{n-1} a r}{S}$ and $\pi^d = \EE\lrb{\ind{\xi^b + \chi < z-1}u}$.
While both $\pi^f$ and $\pi^d$ can be worked out explicitly (see Appendix~\ref{app:pi_prw2}), we next show that their exact expressions are not required to determine $\rho_c$.

Isolating the $\pi^{f/d}$ components of the mean offspring vectors yields the representation
\begin{equation}
    \label{eq:mom_dec}
    M = \overline{M} + \bbd\boldsymbol{\pi}^T
\end{equation}
where $\boldsymbol{\pi} = (0,\pi^f,0, \pi^d)^T$. Since $M$ and $M^T$ have the same spectrum, we look for a critical right eigenvector $\bbv$ of $M^T$, such that
\begin{equation}
    \label{eq:mom_eigven}
    M^T \bbv = \bbv \Leftrightarrow (\overline{M}^T + \boldsymbol{\pi} \bbd ^T)\bbv = \bbv.
\end{equation}
Further imposing that $\bbv$ be orthogonal to $\bbd$ removes the need to compute $\boldsymbol{\pi}$ altogether, at the cost of introducing the overconstrained linear problem

\begin{equation}
    \label{eq:mom_eigvector}
    \overline{M}^T \bbv = \bbv~~\text{and}~~\bbd^T \bbv = 0.
\end{equation}
We show in Appendix~\ref{app:eigvec_prw2} that for $\rho = 1 - (z-1)^{-3} \equiv \rho_c$, such a positive vector $\bbv$ exists, and is explicit.
As a result, $M^T$ and hence $M$ have eigenvalue $1$, which, $\bbv$ being positive, is their Perron root by the Perron--Frobenius theorem \cite{seneta1981nonnegative}, and the critical density for PRW$_2$ is given by
\begin{equation}
    \label{eq:rhocrit_prw2}    
    \rho_c = 1 - \frac{1}{(z-1)^3},
\end{equation}
thus settling the conjecture first formulated in \cite{bonomo2024bethe}.

\subsection{General escape probabilities}

\begin{figure}[h!]
    \includegraphics[width=0.5\textwidth]{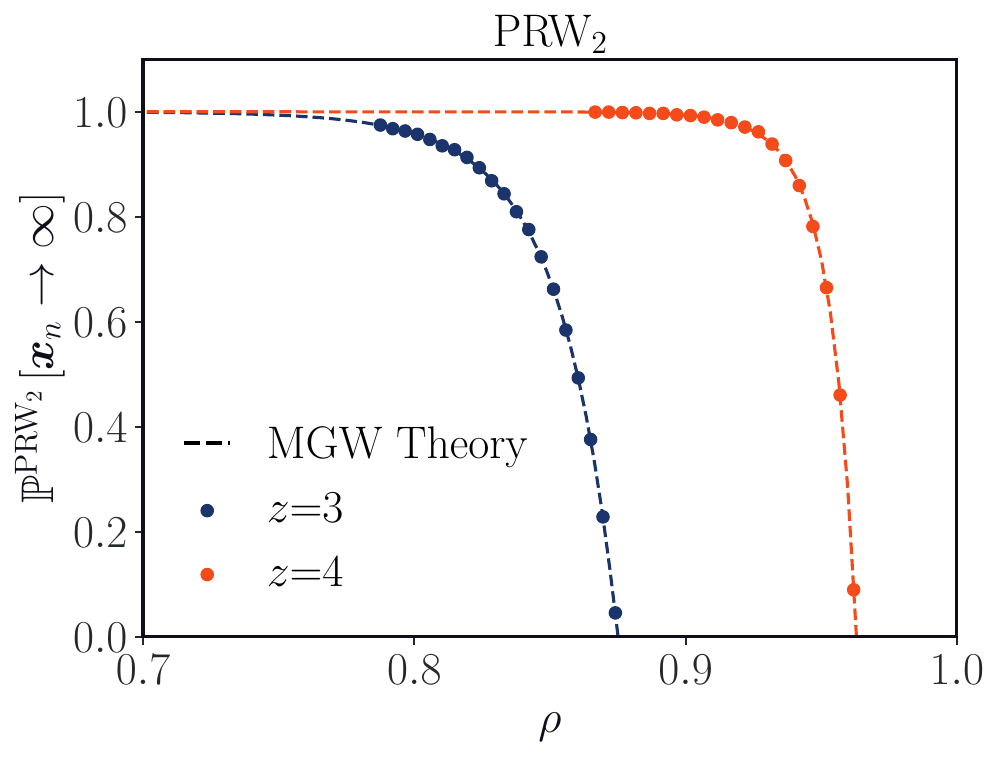}
    \caption{
    Escape probabilities of PRW$_2$ on the Bethe lattice, at varying coordination number $z$. 
    Monte Carlo simulations are averaged over $2\times 10^4$ trials, where each RW is evolved for $10^6$ steps and counts as escaped if the maximum root distance does not saturate, i.e.\ keeps increasing over the last $5\times 10^5$ steps.
    The theoretical dashed lines are obtained by numerically solving the fixed-point equation of the MGW process, with the explicit pgfs of Appendix~\ref{app:pgf_prw2}.
    The critical threshold $\rho_c = 1 - (z-1)^{-3}$ is apparent for both $z=3,4$ values.
    }
    \label{fig:p2_esc}
\end{figure}

We conclude with the analysis of escape probabilities for general $\rho$. As for PRW$_1$, (sub)critical $\rho\geq\rho_c$ immediately implies $\PP^{\text{PRW}_2}\lrb{\bbx_n \to \infty} = 0$, and PRW$_2$ remains bounded almost surely.
For supercritical $\rho < \rho_c$, we determine the pgfs $f^{s,f,b,d}$ of PRW$_2$, as well as the root pgf $f^o$, and provide their explicit expressions in Appendix~\ref{app:pgf_prw2}.
Numerically solving the corresponding fixed-point equation, analogous to Eq.~\eqref{eq:fp_prw1}, then yields the escape probability $\PP^{\text{PRW}_2}\lrb{\bbx_n \to \infty}>0$ for $0<\rho < \rho_c$, which we compare with Monte Carlo simulations of PRW$_2$ in Fig.~\ref{fig:p2_esc}.

\begin{figure}[h!]
    \includegraphics[width=0.48\textwidth]{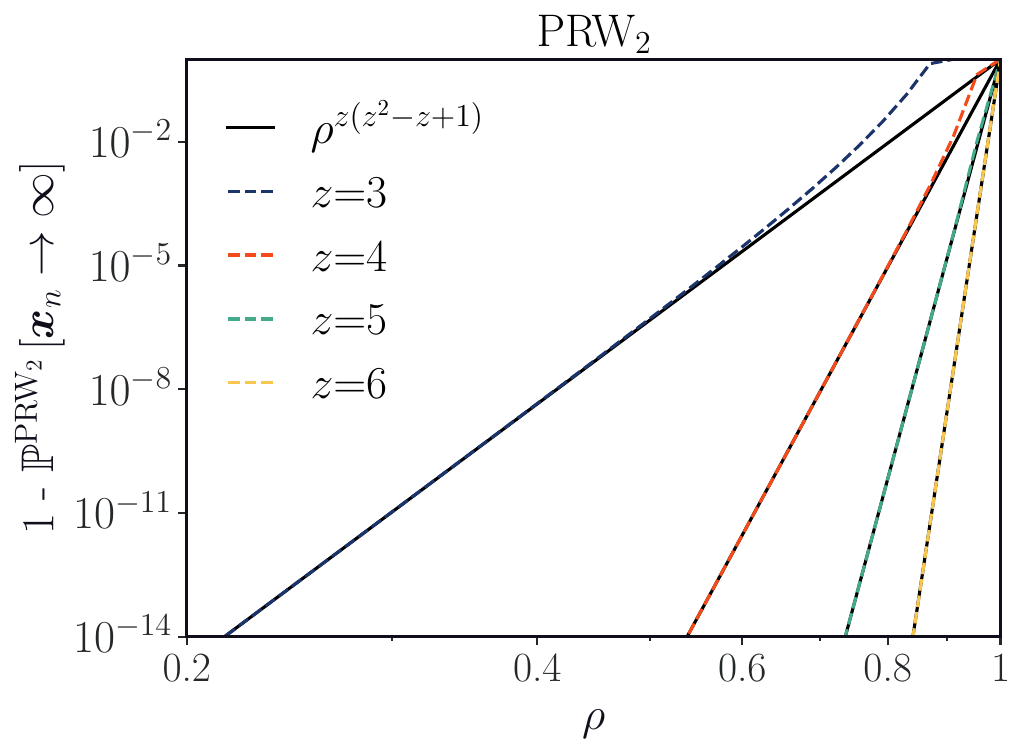}
    \includegraphics[width=0.48\textwidth]{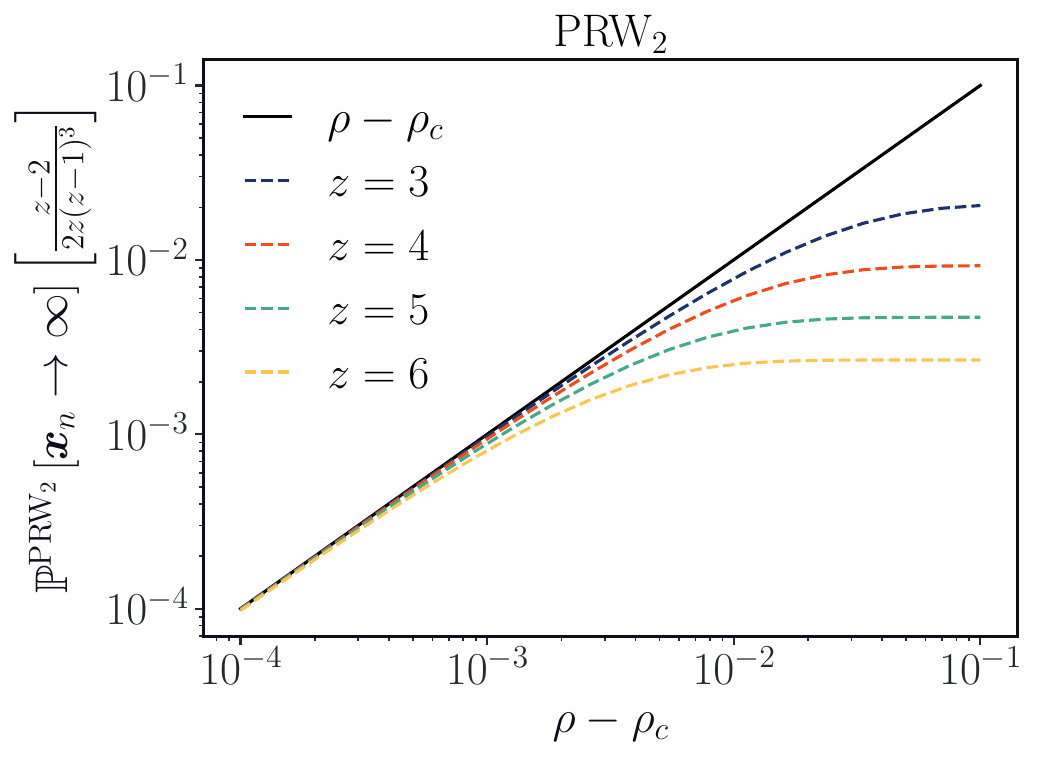}
    \caption{Perturbative behavior of the escape probability of PRW$_2$, across coordination number $z$. In both panels, solid black lines are the first-order predictions and coincide with the dashed lines obtained by numerically solving the fixed-point equation of the MGW process, with the explicit pgfs of Appendix~\ref{app:pgf_prw2}. The curves deviate away from $\rho = 0$ and $\rho = \rho_c$ respectively. \textbf{(left)} Small-$\rho$ behavior of the trapping probability $1-\PP^{\text{PRW}_2}\lrb{\bbx_n\to\infty}$, compared with $\rho^{z(z^2-z+1)}$, Eq.~\eqref{eq:smallrho_prw2}. \textbf{(right)} Close to $\rho_c$, the escape probability is linear in $(\rho_c-\rho)$; once rescaled by the conjectured $z$-dependent slope \eqref{eq:derivative_prw2}, all curves collapse.}
    \label{fig:p2_rho_pert}
\end{figure}

Additionally, the small-$\rho$ behavior is obtained from an expansion of the fixed-point equation (see Appendix~\ref{app:smallrho_prw2}),
\begin{equation}
    \label{eq:smallrho_prw2}
    1 - \PP^{\text{PRW}_2}\lrb{\bbx_n\to \infty} \underset{\rho \to 0}{\sim}\rho^{z(z^2-z+1)},
\end{equation}
as numerically verified in Fig.~\ref{fig:p2_rho_pert} (left).
Finally, we \textit{conjecture} that the behavior close to criticality is given by the simple expression
\begin{equation}
    \label{eq:derivative_prw2}
    \partial_\rho\PP^{\text{PRW}_2}\lrb{\bbx_n\to \infty}|_{\rho_c} = - \frac{2 z (z-1)^3}{z-2}.
\end{equation}
The numerics shown in Fig.~\ref{fig:p2_rho_pert} (right) seem to corroborate this statement, and the recursive structure of the previous escape probability results suggests that it holds, but an exact proof has so far remained out of reach.

\section{Ensemble statistics of PRW traces}
\label{sec:traces}

\begin{figure}[h!]
    \includegraphics[width=0.98\textwidth]{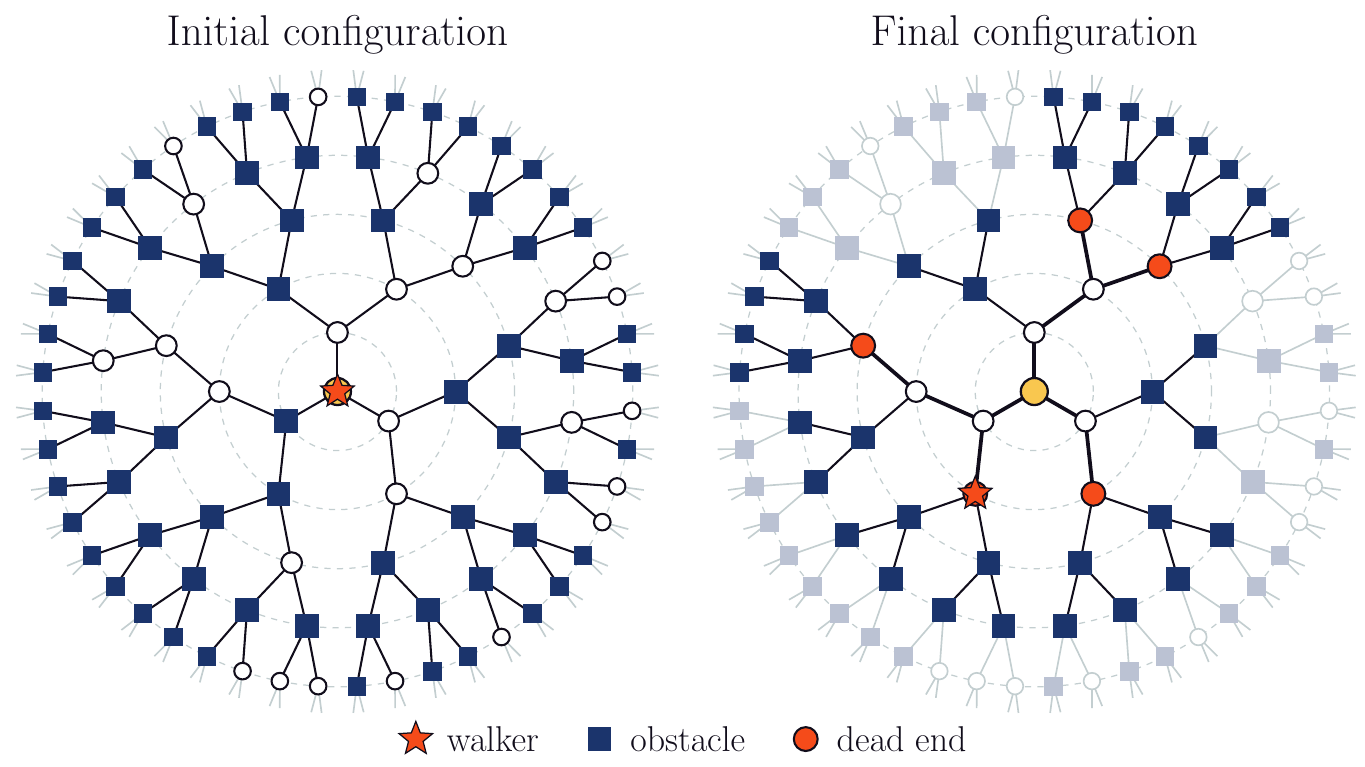}
    \caption{
    Initial and final obstacle configurations for a single PRW$_1$ realization at fixed subcritical density $\rho$. 
    Once no obstacle can be displaced anymore, the walker evolves on a finite graph, comprising 11 nodes, 5 dead ends and with maximum distance to the root equal to 3. 
    }
    \label{fig:trace_schematics}
\end{figure}

It should now be clear that the escape of a pushy random walk on the Bethe lattice is strictly equivalent to the event that the walkable structure generated by repeatedly pushing obstacles is infinite.
In turn, we have shown that these events further coincide with the survival of explicit MGW processes.
In this section, we leverage this construction to obtain refined information on the statistics of the PRW trace, that is, of the set of sites eventually visited by the pushy walker. 
We select a few such observables, and give MGW-type formulas for the maximal depth reached by the PRW, the number of sites in the trace and the number of dead ends in the trace, illustrated in Fig.~\ref{fig:trace_schematics}.
We only consider the (sub)critical case $\rho \geq \rho_c$, for which traces are almost surely finite; the supercritical case $\rho<\rho_c$ 
can be dealt with similarly, provided one conditions on finite traces, at the cost of a notational overload that we do not deem relevant to the overall approach. 
Moreover, we derive our formulas for a generic MGW process with pgfs $\bbf$ and root pgf $f^o$, so that they apply equally to PRW$_1$ and PRW$_2$.

\subsection{Maximum depth of PRW traces}

\begin{figure}[h!]
    \includegraphics[width=0.98\textwidth]{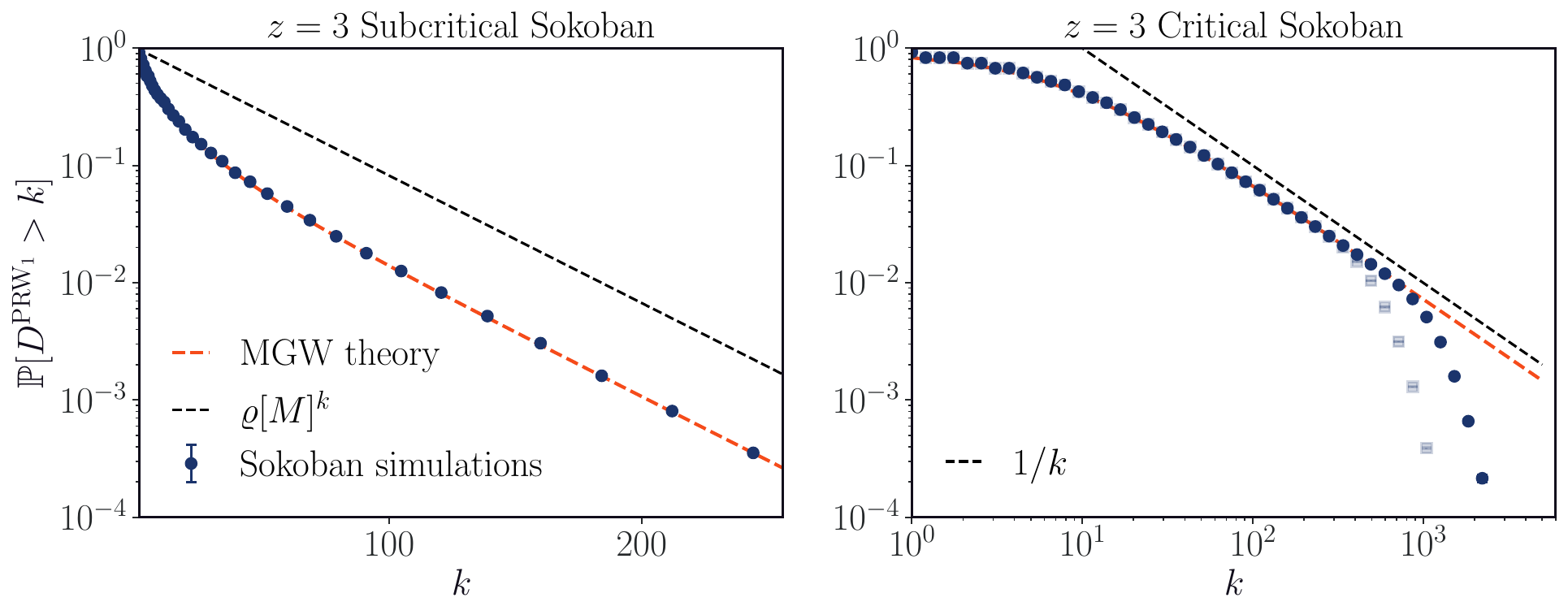}
    \includegraphics[width=0.98\textwidth]{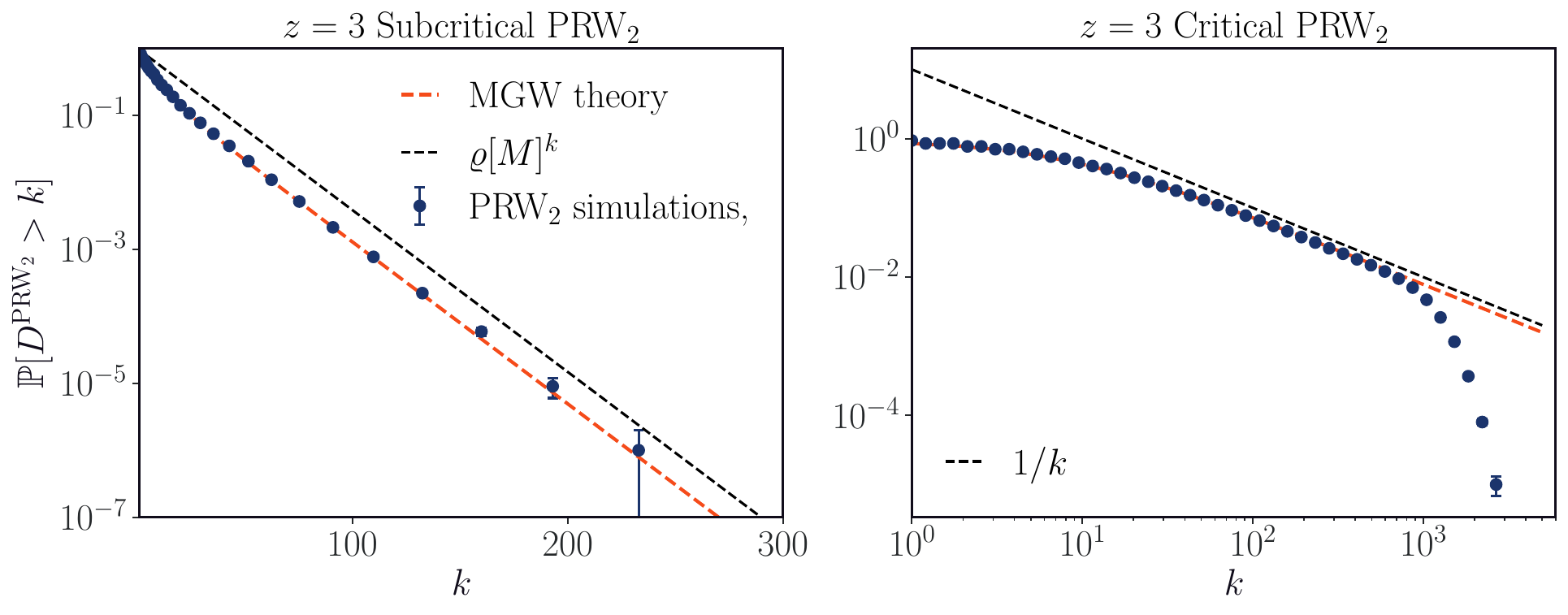}
    \caption{
    Cumulative distribution of $D^{\text{PRW}}$ for both PRW$_1$ {\bf(top row)} and PRW$_2$ {\bf(bottom row)} at $z=3$ and respective subcritical obstacle densities $\rho = 0.7575$ and $\rho=0.88375$ {\bf(left column)} and critical densities $\rho_c = 0.75$ and $\rho_c = 0.875$ {\bf(right column)}.
    The dashed red line is obtained by numerically iterating the extinction probability \eqref{eq:ext_gen} and compared to empirical estimates averaged over $10^7$ (PRW$_1$) and $10^6$ (PRW$_2$) trials capped at a maximum of $10^7$ steps in the subcritical case, and $10^6$ trials capped at $10^8$ steps at criticality.
    {\bf (top right)} At criticality, the algebraic decay of $\PP\lrb{D^{\text{PRW}_{1/2}} > k}$ allows for finite but large traces. In turn, a fair number of trials do not exhaust the walkable structure before reaching the maximum allowed number of steps, which explains the large-$k$ deviation between Monte Carlo estimates and theory. This is verified by performing the same experiments with a lower step cap (shaded points, $10^7$ step cap), and observing that the deviation occurs at lower $k$.   
    }
\label{fig:maxdepth}
\end{figure}

Consider the MGW process $\lrc{\bbZ_k}_{k \in \NN}$ with pgf $\bbf$. As in the previous sections, we denote by $\PP^i$ and $\EE^i$ the probabilities and expectations with respect to the MGW process started from a single node of type $i$.
The notations $\PP^o$ and $\EE^o$ are reserved for the process started from the root, whose offspring distribution may be slightly different (typically, with an extra child).
For $\bbs\in[0,1]^K$, $K$ being the number of types, and an integer vector $\bbZ$, we further introduce the shorthand $\bbs^{\bbZ} = \prod_{i}s_i^{Z^i}$, such that the pgfs explicitly read

\begin{equation}
    \label{eq:pgf_gen}
    f^i(\bbs) = \EE^i\lrb{\bbs^{\bbZ_1}}.
\end{equation}
Recalling that $|\bbZ_k| = \sum_j Z_k^j$ denotes the total number of nodes at generation $k$, the probabilities $q^i_k = \PP^i\lrb{|\bbZ_k| = 0}$ that the process has died out by generation $k$ satisfy the vector recurrence relation
\begin{equation}
    \label{eq:ext_gen}
    \bbq_{k+1} = \bbf(\bbq_k) = \bbf^{(k+1)}(\bbq_0),
\end{equation}
where $\bbf^{(k)}$ denotes the $k$-fold composition of $\bbf$, $\bbq_0 = \boldsymbol{0}$, and $\bbq_k$ converges to $\bbq = \boldsymbol{1}$ whenever $\rho \geq \rho_c$.
Finally, for the MGW process started from the root, one has, for $k\geq 1$,
\begin{equation}
    \label{eq:ext_root_gen}
    q_k^o = \PP^o\lrb{|\bbZ_k| = 0} = f^o(\bbq_{k-1}).
\end{equation}
In turn, the distribution of the maximum depth $D$ of the MGW process, that is, the depth of its deepest node, is directly obtained from $q^o_k$ as
\begin{equation}
    \label{eq:maxdepth_root_gen}
    \PP^o\lrb{D \leq k} = \PP^o\lrb{|\bbZ_{k+1}| =  0} \implies \PP^o\lrb{D = k} = q_{k+1}^{o} - q_{k}^{o}.
\end{equation}
As a result, iterating the extinction recurrence \eqref{eq:ext_gen} yields the full distribution of $D$.
To relate $D$ to the maximum depth reachable by PRW, we recall that the leaves of the MGW process, that is, \textit{childless nodes}, correspond to obstacles that cannot be displaced further away from the root. Since PRW$_k$ can never enter such nodes, and every node that PRW can enter has $z-1$ children, the deepest generation of a finite MGW tree only contains childless nodes. As a result, the maximum depth reachable by PRW is simply
\begin{equation}
    D^{\text{PRW}} = D-1.
\end{equation}
Note that, by definition of the root, $f^o(\boldsymbol{0}) = 0$, such that $\PP^o\lrb{D = 0} = 0$, which guarantees that $D^{\text{PRW}}$ is always non-negative. The agreement between Eq.~\eqref{eq:maxdepth_root_gen} and Monte Carlo simulations of both PRW$_1$ and PRW$_2$ is shown in Fig.~\ref{fig:maxdepth}.

We highlight that the large-$k$ decay of the distribution of $D^{\text{PRW}}$ simply follows from known convergence results on $\bbq_k$.
For strict subcriticality $\rho > \rho_c$, $\bbq_k$ converges exponentially fast to $\boldsymbol{1}$ \cite{athreya1972branching}, with rate given by the Perron root $\varrho\lr{M}<1$ of the mean offspring matrix $M$, so that
\begin{equation}
    \label{eq:conv_exp}
    \PP^o\lrb{D^{\text{PRW}} > k} \underset{k\to \infty}{\propto} \varrho\lr{M}^{k}.
\end{equation}
At the critical density $\rho = \rho_c$, $\bbq_k$ still converges to $\boldsymbol{1}$, but much more slowly, and the distribution of $D^{\text{PRW}}$ decays as a power law \cite{kesten1966galton,joffe1967multitype},
\begin{equation}
    \label{eq:conv_alg}
    \PP^o\lrb{D^{\text{PRW}} > k} \underset{k\to \infty}{\propto} k^{-1}.
\end{equation}
Both asymptotic behaviors are recovered in the actual simulations of PRW$_{1/2}$ shown in Fig.~\ref{fig:maxdepth}.

\subsection{Volume of PRW traces}

Since the walkable structure is almost surely finite, PRW eventually visits every accessible node, and we denote by $\mathcal{N}^o$ the total number of visited nodes, i.e.\ the volume of the trace.
More generally, it is useful to define the total node number $\mathcal{N}^i$ of an MGW tree started from a node of type $i$.
Leveraging the tree structure of the MGW process, and the fact that a childless node can never be accessed by PRW, we obtain the following \textit{recursive distributional equation} \cite{aldous2005survey}

\begin{equation}
    \label{eq:rde_vol}
    \mathcal{N}^i = \ind{|\bbn| \geq 1}\lr{1 + \sum_{t=1}^{K}\sum_{i_t=1}^{n_t}\mathcal{N}^{t, (i_t)}},
\end{equation}
where the type counts $\bbn = (n_1, \dots, n_K)$ are drawn from the offspring distribution of a type-$i$ node, $\bbn\sim\bbX^i$, and, for each type $t$, the random variables $\mathcal{N}^{t, (i_t)}$ are independent copies of $\mathcal{N}^t$.
While the random variables on the right-hand side are mutually independent (conditionally on their types), they are strongly correlated with the random variable on the left-hand side.
As a result, the volume statistics are most easily understood through their pgfs $\mathcal{N}^i(s) = \EE^i\lrb{s ^{\mathcal{N}^i}}$.
From Eq.~\eqref{eq:rde_vol}, a direct computation shows that these pgfs satisfy the explicit vector functional equation (reminescent of the classical (M)GW approaches first investigated in \cite{otter1949multiplicative,good1960generalizations})

\begin{equation}
    \label{eq:rde_vol_pgf}
    \boldsymbol{\mathcal{N}}(s) = s\,\bbf\lr{\boldsymbol{\mathcal{N}}(s)} + (1-s)\bbp_0,~~~\mathcal{N}^o(s) = s\,f^o\lr{\boldsymbol{\mathcal{N}}(s)},
\end{equation}
where $\bbp_0 = \bbf(\boldsymbol{0})$ collects the probabilities $p_0^i$ that a type-$i$ node has no children, and where we have isolated the special role played by the root.
It is clear from Eq.~\eqref{eq:rde_vol_pgf} that $\boldsymbol{\mathcal{N}}(0)$, the probability that a node is inaccessible to PRW, is equal to $\bbp_0$, the probability that the node has no children, as required by the physical constraints underlying PRW$_k$'s dynamics.
Notice, however, that $\mathcal{N}^o(0) = 0$, since the root is always accessible to PRW.

The functional equation \eqref{eq:rde_vol_pgf} uniquely determines $\mathcal{N}^o(s)$, or, equivalently, the distribution $\PP^o\lrb{\mathcal{N}^o = k}$.
Although analytical expressions for the latter are typically out of reach, the small number of types (two for PRW$_1$, four for PRW$_2$) allows for efficient numerical evaluations of $\PP^o\lrb{\mathcal{N}^o = k}$.
Formally, pgf inversion via Cauchy's residue formula \cite{flajolet2009analytic} yields
\begin{equation}
    \label{eq:pgf_inversion}
    \PP^o\lrb{\mathcal{N}^o = k} = \frac{1}{2\pi i}\oint \frac{\mathcal{N}^o(w)}{w^{k+1}}\der w,
\end{equation}
where the complex integral is taken along a closed contour around the origin. Discretizing Eq.~\eqref{eq:pgf_inversion} yields an approximation of $\PP^o\lrb{\mathcal{N}^o = k}$ involving only a finite number of values $\lrc{\mathcal{N}^o(w_j)}_{0 \leq j < J}$, which are in turn obtained by numerically solving the fixed-point equation \eqref{eq:rde_vol_pgf} at each $w_j$.
These numerical strategies (see Appendix~\ref{app:numerics} for implementation details) provide arbitrarily precise numerical approximations of $\PP^o\lrb{\mathcal{N}^o = k}$, which seamlessly agree with Monte Carlo simulations of PRW$_{1/2}$ shown in Fig.~\ref{fig:trace_stat}.

We end this subsection with a comment on the large-$k$ tail of the volume distribution.
One should not expect the large-$k$ decay of $\PP^o\lrb{\mathcal{N}^o = k}$ to be identical to that of $D^{\text{PRW}}$, since the nodes belonging to the longest path from the root form only a small fraction of the total walkable structure%
.
The determination of the functional form of this decay (exponential with subexponential corrections) is a classical topic in generating function theory \cite{flajolet2009analytic}.
Even for single-type GW processes, the decay rate is typically only defined implicitly, via yet another functional equation, and must be determined numerically.
We expect that a corresponding multi-type argument can be adapted from the literature \cite{good1960generalizations}, but leave this analysis for future work.

\subsection{Dead ends of PRW traces}

\begin{figure}[h!]
    \includegraphics[width=0.48\textwidth]{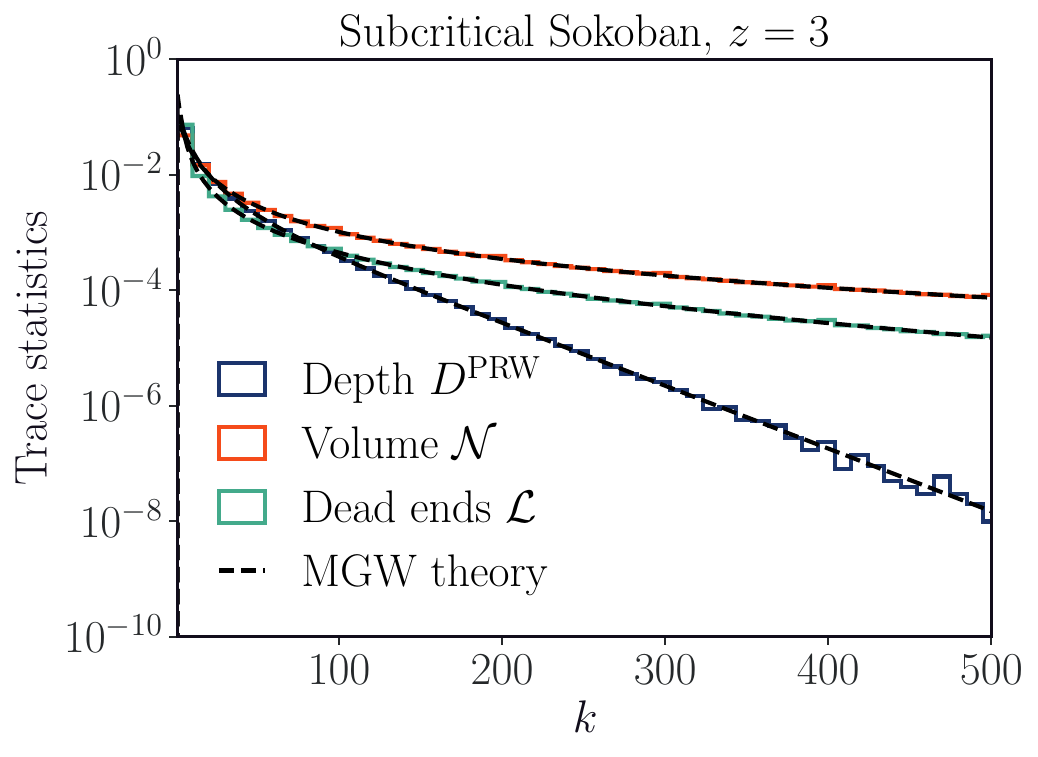}
    \includegraphics[width=0.48\textwidth]{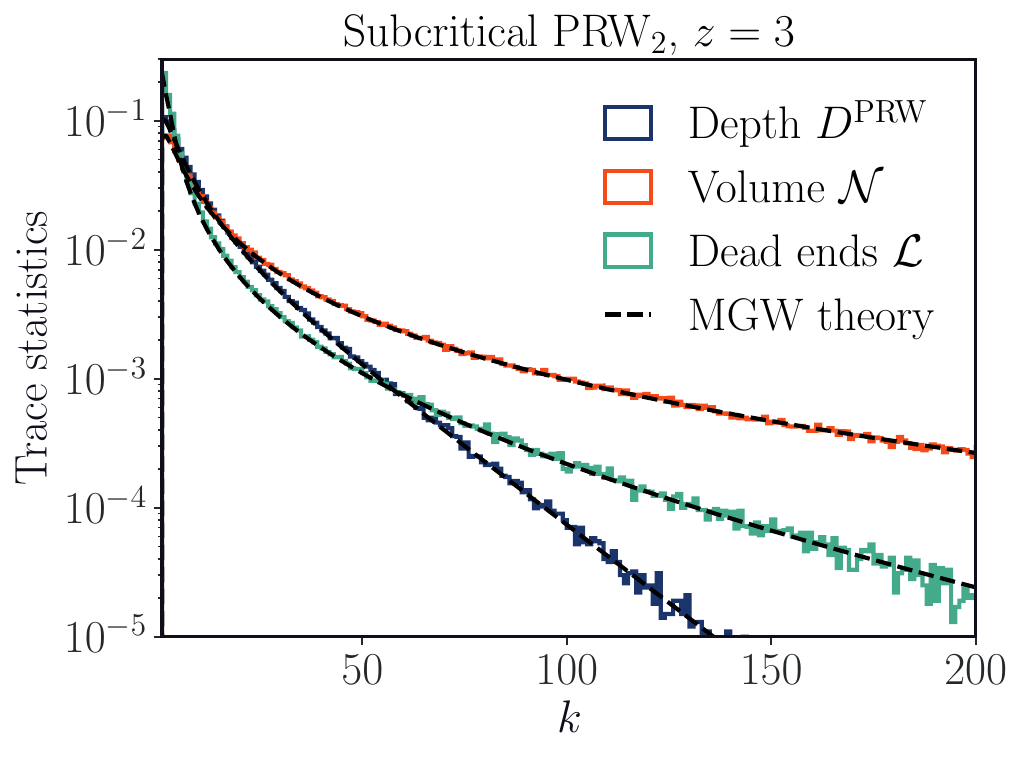}
    \caption{Empirical distributions of maximum depth, volume and number of dead ends for subcritical PRW$_1$ {\bf (left)} and PRW$_2$ {\bf (right)} at respective densities $\rho = 0.7575$ and $\rho = 0.88375$.
    The theoretical dashed lines obtained by approximate inversion of the corresponding generating functions are compared to $10^7$ (PRW$_1$) and $10^6$ (PRW$_2$) independent trials, capped at $10^7$ maximum steps.
    Note that while $D^{\text{PRW}}$ displays clear exponential decay, this is not the case for $\mathcal{L}$ and $\mathcal{N}$, for which we expect subexponential prefactors.
    }
    \label{fig:trace_stat}
\end{figure}

We finally consider the number of dead ends $\mathcal{L}^o$ in the PRW trace.
Similarly to $\mathcal{N}$, we introduce the typed versions $\mathcal{L}^i$, and aim to derive the functional vector equation satisfied by the generating functions $\mathcal{L}^i(s)= \EE^i\lrb{s^{\mathcal{L}^i}}$.
A node is a dead end for PRW if and only if it is accessible (that is, it has children), but none of its children is accessible. For a type-$i$ node, we denote by $\mathcal{E}^i$ this event, and by $\ell^i$ the corresponding probability, which explicitly reads $\ell^i = f^i(\bbp_0) - p_0^i$. In turn, one obtains the recursive distributional equation

\begin{equation}
    \label{eq:rde_de}
    \mathcal{L}^i = \ind{\mathcal{E}^i} + (1 - \ind{\mathcal{E}^i})\sum_{t=1}^{K}\sum_{i_t=1}^{n_t}\mathcal{L}^{t, (i_t)},
\end{equation}
with the same notations as in Eq.~\eqref{eq:rde_vol}.
Note that the above recursion implies that a node with no children does not contribute to the dead ends of PRW, since it is not accessible. Conversely, any accessible node with a finite subtree has at least one dead end below it. As a result, $\PP^i\lrb{\mathcal{L}^i = 0} = p_0^i$.
Furthermore, $\mathcal{L}^i \geq 2$ can only occur when the source node has at least one accessible child, in which case the following equality holds:

\begin{equation}
    \label{eq:rde_de_2}
    \forall k \geq 2 :~~~~ \PP^i\lrb{\mathcal{L}^i=k} = \PP^i\lrb{\sum_{t=1}^{K}\sum_{i_t=1}^{n_t}\mathcal{L}^{t, (i_t)} = k}.
\end{equation}
The case $k=1$ is trickier: either the source node has no accessible children and is itself the dead end, or it has exactly one accessible child, whose subtree also has exactly one dead end. 
This observation is formalized with the probabilistic identity 

\begin{equation}
    \label{eq:rde_de_3}
    \PP^i\lrb{\mathcal{L}^i=1} = \ell^i + \sum_j \partial_{s_j} f^i(\bbp_0)\,\PP^{j}\lrb{\mathcal{L}^j = 1},
\end{equation}
and together Eqs.~\eqref{eq:rde_de}--\eqref{eq:rde_de_3} yield the functional equation satisfied by $\boldsymbol{\mathcal{L}}(s)$,

\begin{equation}
    \label{eq:rde_de_pgf}
    \boldsymbol{\mathcal{L}}(s) = \bbf\lr{\boldsymbol{\mathcal{L}}(s)} + (s-1)\boldsymbol{\ell},~~~\mathcal{L}^o(s) = f^o\lr{\boldsymbol{\mathcal{L}}(s)} + (s-1)\ell^o,
\end{equation}
where $\boldsymbol{\ell} = (\ell^i)_i$ and $\ell^o = f^o(\bbp_0)$. Note that we explicitly recover $\boldsymbol{\mathcal{L}}(0) = \bbp_0$ and $\mathcal{L}^o(0) = 0$, so that the PRW trace always has at least one dead end, potentially the root itself.
Similarly to volume statistics, the functional equation \eqref{eq:rde_de_pgf} is amenable to approximate fixed-point solutions and pgf inversion (see Appendix~\ref{app:numerics} for details); 
the resulting dead-end statistics display satisfactory agreement with the Monte Carlo simulations of PRW$_1$ and PRW$_2$ shown in Fig.~\ref{fig:trace_stat}.

\section{Conclusion}
\label{sec:conclusion}

Pushy random walks offer a novel microscopic paradigm to capture complex and intriguing transport phenomena arising from particle--environment interactions.
In particular, the loss of the percolation transition on $\ZZ^2$ \cite{bonomo2023loss} -- which has yet to be formally proven -- has motivated the analysis of Sokoban dynamics (PRW$_1$) on the traditionally tractable Bethe lattice \cite{bonomo2024bethe}, on which the percolation transition has been shown to survive.
In this article, we revisit pushy dynamics on the Bethe lattice and show that the tree structure of the substrate allows for a drastic simplification: the random system comprising the walker and the displaceable obstacles is statistically equivalent to a walker evolving on a static random tree, as in a quenched Lorentz gas, albeit with correlated, multi-type statistics.

Specifically, this effective static structure is constructed by introducing explicit multi-type Galton--Watson processes, with two types for the Sokoban walk and four types for the second-order pushy random walk.
This mapping provides a straightforward way to recover and extend the analysis of the escape properties of the Sokoban walk, as well as to tackle the previously unexplored PRW$_2$ case.
In particular, we compute the critical densities $\rho_c$, establishing $\rho_c = 1-(z-1)^{-3}$ for PRW$_2$ as conjectured in \cite{bonomo2024bethe}, analyze the asymptotic behavior close to criticality, and propose numerical routes to compute escape probabilities arbitrarily far from criticality.
Furthermore, we find that the Galton--Watson framework is naturally suited to probe the statistical properties of the dynamical structure -- or trace -- generated by pushy random walkers as they explore the lattice and rearrange obstacles.
In fact, we put forward a systematic methodology to obtain the distributions of important geometric features, such as the maximum distance from the initial point, the total explored volume, and the total number of dead ends.

This approach opens up a wealth of potential extensions and new questions, which we plan to investigate in future works.
First, as alluded to in the main text, investigating higher-order pushy walkers requires identifying the corresponding MGW processes, with an increasing number of types.
While this might be combinatorially tedious, the methodology remains unchanged.
Note, however, that the simple recursive structure of both $\rho_c$ and the critical derivative $\partial_\rho \PP^{\text{PRW}_k}\lrb{\bbx_n\to\infty}|_{\rho_c}$ suggests a simpler route, or a recursive construction in $k$, which remains to be identified.
Next, the GW mapping we exhibit and exploit throughout is expected to hold for a broad class of tree-like structures.
In particular, paradigmatic tree-like transport substrates include fractal T-graphs, spanning trees, comb-like structures, or even GW trees themselves.
We expect most of the results derived in this paper to translate naturally to such diverse scenarios, with only slight modifications.
Finally -- and most importantly -- we have so far only dealt with \textit{static} properties: escape probabilities and geometrical features of the explored territory.
However, PRW$_k$ are inherently \textit{dynamical} objects, and a number of questions remain to be addressed, among which cover times and escape speeds.
We emphasize that the GW mapping remains particularly relevant in this case, and expect that the vast literature on random walks on GW trees \cite{lyons1995ergodic,lyons1996biased,aidekon2014speed,dembo2012clt} will yield concrete answers to these questions.

\section*{Acknowledgments}

The author warmly thanks S. Reuveni on two counts: first, for introducing the author to the Sokoban model and discussing its open questions during the September 2026 \textit{Art of Randomness} conference -- organized by C. Le Vaou, A. Rosso and G. Schehr in Paris in honor of S. Majumdar's many contributions -- and, second, for his insights and advice during the preparation of this manuscript.
The author, as the recipient of the Postdoctoral Laplace Chair, acknowledges funding from Capital Fund Management (CFM), which made this independent work possible.

\section*{LLM Usage Disclaimer}

We acknowledge the use of LLMs to assist grammar, spelling and (some) wording clarifications. 
The tool was also used to support the writing of the code (Monte Carlo simulations, fixed point solves and pgf inversion) as well as drawing some schematics.
AI-generated content was rigorously reviewed, verified, and edited by the author. The author bear full responsibility for the originality and scientific integrity of this work.

\bibliography{refs}

\newpage
\appendix

\section{Exact formulas for the $z=3$ case}
\label{app:z3}

For $z=3$, the fixed-point equations of the Lorentz gas and of PRW$_1$ are of low enough degree to be solved in closed form.

For the Lorentz gas, Eq.~\eqref{eq:fixed_point_gw} with the pgf \eqref{eq:pgf_lorentz} reduces to the quadratic equation
\begin{equation}
    \label{eq:z3_lorentz}
    (1-\rho)^2 q^2 - \lrb{1-2\rho(1-\rho)}q + \rho^2 = 0,
\end{equation}
whose roots are $1$ and $\rho^2/(1-\rho)^2$, hence $q = \min\lr{1, \rho^2/(1-\rho)^2}$.
Below the threshold $\rho_c = 1/2$, the fixed-point equation then implies $\rho + (1-\rho)q = \sqrt{q} = \rho/(1-\rho)$, so that the escape probability \eqref{eq:esc_lorentz_final} is fully explicit,
\begin{equation}
    \label{eq:esc_lorentz_z3}
    \PP^{\text{PRW}_0}\lrb{\bbx_n \to \infty} = 1 - \lr{\frac{\rho}{1-\rho}}^3, ~~~~ \rho < \rho_c = \frac{1}{2},
\end{equation}
and vanishes for $\rho \geq 1/2$.

For PRW$_1$, the fixed-point equation \eqref{eq:fp_prw1} with the pgfs \eqref{eq:exp_pgf_prw1} reduces to a cubic equation, whose relevant root for $\rho<\rho_c$ is
\begin{equation}
    \label{eq:z3_prw1}
    \begin{split}
        q^m = \frac{1}{6} \Bigg[~~&\frac{4 \sqrt[3]{2} (\rho -1) (2 \rho +1)}{\sqrt[3]{(\rho -1)^4 \left(4 \rho  (4 \rho +1)+3
   \sqrt{24 \rho  (2 \rho +1)+9}+7\right)}} ~ + \\
    &\frac{2^{2/3} \sqrt[3]{(\rho -1)^4 \left(4 \rho  (4 \rho +1)+3
   \sqrt{24 \rho  (2 \rho +1)+9}+7\right)}}{(\rho -1)^2} - 2 \Bigg], 
    \end{split}
\end{equation}
and $\bbq = ((q^m)^3, q^m)$. The corresponding $\PP^{\text{PRW}_1}\lrb{\bbx_n \to \infty}$ is obtained by plugging $\bbq$ into $f^o$. 

\section{Schematics for the displacement rules of PRW$_2$}
\label{app:rules_prw2}

\begin{figure}[h!]
    \includegraphics[width=\textwidth]{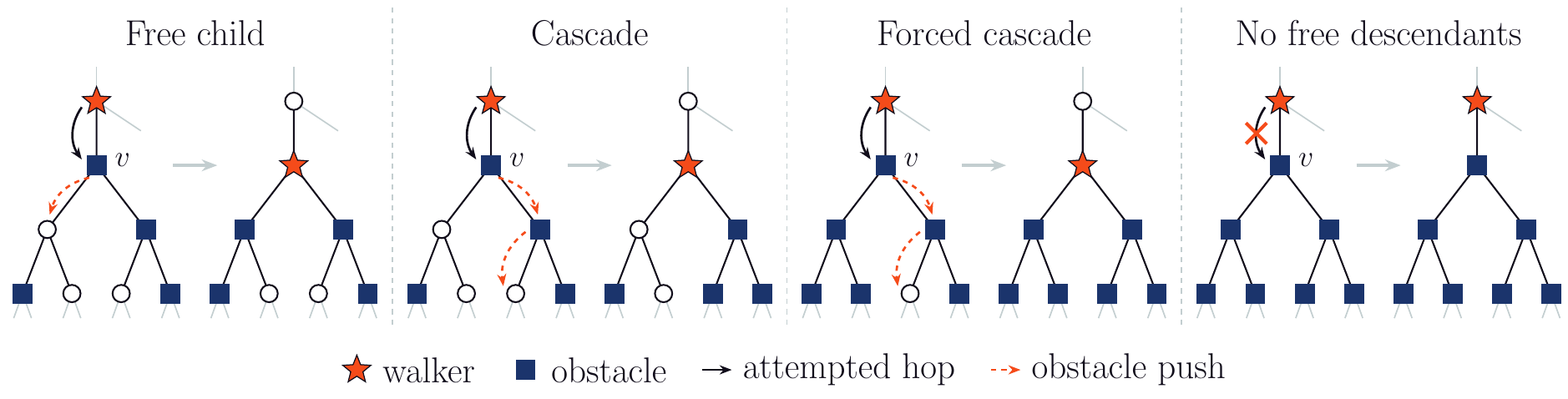}
    \caption{
        }    
    \label{fig:rules_prw2}
\end{figure}

Figure~\ref{fig:rules_prw2} illustrates the displacement rules of PRW$_2$ stated in Section~\ref{sec:prw2}, for $z=3$, in four representative situations arising when the walker attempts to hop onto an obstructed node $v$.
In the first two panels (\textit{Free child} and \textit{Cascade}), $v$ has two admissible children: a free one, and an obstructed one with a free child of its own. Each of them is chosen with probability $1/2$, so that the obstacle on $v$ either occupies the free child, or cascades through the obstructed child, whose obstacle is pushed onto its free child.
In the third panel (\textit{Forced cascade}), both children of $v$ are obstructed and only one of them has a free child, so that the cascade occurs with probability $1$.
In the last panel (\textit{No free descendants}), all the descendants of $v$ in the two layers below it are obstructed, and the hop fails.
In the first three cases, the walker moves to $v$, and only the subtree hanging from $v$ is modified.

\section{Mean offspring matrix of PRW$_2$}
\label{app:mom_prw2}

We use the notations of Section~\ref{sec:prw2}, $(n,a,b,c,r,S) = (z-1,\ 1-\rho,\ \rho(1-\rho^{z-1}),\ \rho\rho^{z-1},\ \rho^{z-1},\ 1-\rho^{z-1})$, and order types as $(s,f,b,d)$. With the convention $M_{ij} = \EE\lrb{X^j_i}$, the columns of $M$ are the mean offspring vectors \eqref{eq:mov1_prw2}--\eqref{eq:movd_prw2}, i.e.
\begin{equation}
    \label{eq:app_mom_prw2}
    M = \lr{\EE\lrb{\bbX^s},~\EE\lrb{\bbX^f},~\EE\lrb{\bbX^b},~\EE\lrb{\bbX^d}} = \overline{M} + \bbd\boldsymbol{\pi}^T,
\end{equation}
with $\bbd = (-1,2-r,r,-1)^T$ and $\boldsymbol{\pi} = (0,\pi^f,0,\pi^d)^T$.
The $f$- and $d$-type vectors involve the moments of the multinomial law of $\xi = (\xi^s,\xi^f,\xi^b)^T$ conditioned on $\xi^s>0$, whose probability is $S = 1-\rho^{n}$,
\begin{equation}
    \label{eq:app_cond_moments}
    \begin{split}
        &e_s \equiv \EE\lrb{\xi^s|\xi^s>0} = \frac{na}{S},~~~~e_f \equiv \EE\lrb{\xi^f|\xi^s>0} = \frac{nb(1-\rho^{n-1})}{S},\\
        &e_b \equiv \EE\lrb{\xi^b|\xi^s>0} = \frac{nc(1-\rho^{n-1})}{S},
    \end{split}
\end{equation}
where, for instance, $\EE\lrb{\xi^f\ind{\xi^s>0}} = nb - nb\rho^{n-1}$, since, given that one child is of type $f$, $\xi^s = 0$ requires its $n-1$ siblings to be occupied. Similarly, the $d$-type node is a dead end when $\xi = (1,0,n-1)$ and its single free child received a $b$-type obstacle, $\chi=1$, which yields $p = nac^{n-1}r/S$.
Rows being indexed by the parent type, the $\pi$-free part of the mean offspring matrix finally reads
\begin{equation}
    \label{eq:app_mbar_prw2}
    \overline{M}^T = \begin{pmatrix}
        na & nb & nc & 0\\
        e_s & e_f - 1 & e_b & 1\\
        0 & n(1-r) - (1-r^n) & n(r-r^n) & 1-r^n\\
        e_s - 1 & e_f - r + p & e_b + r - np & 1-p
    \end{pmatrix}.
\end{equation}

\subsection{Pushing probabilities $\pi^f$ and $\pi^d$}
\label{app:pi_prw2}

The probabilities that the obstacle of an $f$- or $d$-type node is pushed onto a free child follow directly from the laws \eqref{eq:rvf_prw2} and \eqref{eq:rvd_prw2},
\begin{equation}
    \label{eq:app_pi_prw2}
    \begin{split}
        \pi^f &= \EE\lrb{\frac{\xi^s}{\xi^s+\xi^f}\Big|\xi^s>0},\\
        \pi^d &= \EE\lrb{\ind{\xi^b+\chi<z-1}\frac{\xi^s-1}{\xi^s-1+\xi^f+1-\chi}\Big|\xi^s>0},
    \end{split}
\end{equation}
where the expectations run over the conditioned multinomial law of $\xi = (\xi^s,\xi^f,\xi^b)^T$ and over $\chi\sim\text{Ber}(r)$.
As in Appendix~\ref{app:pgf_prw2}, writing $\xi^s/(\xi^s+\xi^f) = \int_0^1\xi^s t^{\xi^s+\xi^f-1}\,\der t$ turns $\pi^f$ into an elementary integral, and, using $a+b+c = 1$,
\begin{equation}
    \label{eq:app_pif_prw2}
    \pi^f = \frac{1}{S}\int_0^1 na\lrb{(a+b)t+c}^{n-1}\der t = \frac{a\lr{1-c^{n}}}{(1-c)(1-r)}.
\end{equation}
The extra conditioning and the shifted counts prevent a similar simplification for $\pi^d$, which only admits the finite-sum representation
\begin{equation}
    \label{eq:app_pid_prw2}
    \pi^d = \frac{1}{S}\sum_{j=2}^{n}\sum_{m=0}^{n-j}\binom{n}{j}\binom{n-j}{m}(j-1)\,a^jb^mc^{n-j-m}\lrb{\frac{1-r}{j+m} + \frac{r}{j+m-1}}.
\end{equation}

\section{Explicit critical eigenvector of $\overline{M}^T$}
\label{app:eigvec_prw2}

We look for a vector $\bbv = (1, v_f, v_b, v_d)^T$, normalized such that $v_s = 1$, solving the overconstrained problem \eqref{eq:mom_eigvector}, i.e.\ $\overline{M}^T\bbv = \bbv$ and $\bbd^T\bbv = 0$.
We first solve the $f$ and $b$ rows of $\overline{M}^T\bbv = \bbv$, together with the orthogonality condition $\bbd^T\bbv = 0$. Treating $\rho$, $r = \rho^n$ and $R = r^n = \rho^{n^2}$ as independent symbols, this linear system has the explicit solution, valid for arbitrary $(\rho,z)$,
\begin{equation}
    \label{eq:app_v_general}
    \begin{split}
        &v_f = \frac{R\lrb{n^2(1-\rho) - n(1-r)^2} - n(1-\rho)(nr-1) - (1-r)}{(1-r)\Delta},\\
        &v_b = \frac{R(nr-n+1) - n^2\rho + n^2 - 1}{\Delta},~~~~v_d = (2-r)v_f + rv_b - 1,\\
        &\Delta = R\lr{n^2r - n^2\rho + nr + r} - n\rho.
    \end{split}
\end{equation}
Plugging this solution into the two remaining rows, the residuals read
\begin{equation}
    \label{eq:app_v_residuals}
    \lr{\overline{M}^T\bbv - \bbv}_s = -\frac{Rr}{\Delta}\lr{n^3\rho - n^3 + 1},~~~~\lr{\overline{M}^T\bbv - \bbv}_d = \frac{Rnr(1-\rho)}{\rho(1-r)\Delta}\lr{n^3\rho - n^3 + 1},
\end{equation}
and both vanish if and only if $n^3\rho - n^3 + 1 = 0$, that is, at $\rho = \rho_c = 1 - (z-1)^{-3}$.
At this density, the dependence on $R$ drops out, and the critical eigenvector takes the simple form
\begin{equation}
    \label{eq:app_v_crit}
    \bbv = \frac{1}{(n^2+n+1)(1-r)}\begin{pmatrix}
        (n^2+n+1)(1-r)\\ 1+n(1-r)\\ n(1-r)\\ 1 - n(n-1)(1-r)
    \end{pmatrix},~~~~r = \rho_c^{\,n} = \lr{1 - \frac{1}{n^3}}^n.
\end{equation}
Finally, one checks that $\bbv$ is positive. 
Since $r \geq 1 - n^{-2}$, we have $0 < 1-r\leq n^{-2}$, hence $n(n-1)(1-r)\leq (n-1)/n < 1$, and all components of Eq.~\eqref{eq:app_v_crit} are positive.
\paragraph{Uniqueness of $\rho_c$.} The argument above shows that $\varrho(M) = 1$ at $\rho = \rho_c$, but does not exclude other densities at which $\varrho(M) = 1$, with $\bbd^T\bbv\neq 0$. We do not prove uniqueness here, but it is physically clear that adding obstacles can only make it harder for the walker to escape, so that the extinction probability $q^o$ is a nondecreasing function of $\rho$. Assume then that the escape probability vanishes for some $\rho<\rho_c$. By monotonicity, it would vanish on a whole neighborhood of $\rho_c$, and its derivative at $\rho_c$ would be zero. For PRW$_1$, this is ruled out by the nonzero derivative \eqref{eq:derivative_prw1}. For PRW$_2$, the conjectured derivative \eqref{eq:derivative_prw2}, supported by numerics, is nonzero as well, and we expect $\rho_c = 1-(z-1)^{-3}$ to be the unique critical density.

\section{Probability generating functions of PRW$_2$}
\label{app:pgf_prw2}

With $\bbs = (s_1,s_2,s_3,s_4)$ conjugate to the types $(s,f,b,d)$, the $s$-type and root pgfs follow from the multinomial laws,
\begin{equation}
    \label{eq:app_pgf_s_prw2}
    f^s(\bbs) = \lr{a s_1 + b s_2 + c s_3}^{z-1}, ~~~~ f^o(\bbs) = \lr{a s_1 + b s_2 + c s_3}^{z},
\end{equation}
where $g(\bbs) = a s_1 + b s_2 + c s_3$ is the pgf of the type of a child with untouched Lorentz statistics. We also introduce $h(\bbs) = (1-r)s_2 + r s_3$, the pgf of the type of an occupied node with untouched children, which is of type $f$ or $b$ with probability $1-r$ or $r$, respectively.

The three remaining types share a common structure. Upon entering a node of type $i\in\{f,b,d\}$, one first reveals the configuration $\nu = (\nu^s,\nu^f,\nu^b)$ of its free, $f$-type and $b$-type children, drawn from a type-dependent law. The obstacle is then pushed either onto a free child, with probability $\nu^s/(\nu^s+\nu^f)$, which becomes of type $f$ or $b$, or onto an $f$-type child, with probability $\nu^f/(\nu^s+\nu^f)$, which becomes of type $d$. If $\nu^s+\nu^f = 0$, the node is a dead end and has no offspring.
From Eqs.~\eqref{eq:rvb_prw2}--\eqref{eq:rvd_prw2}, the generating functions $G^i(x,y,w) = \EE\lrb{x^{\nu^s}y^{\nu^f}w^{\nu^b}}$ of these configurations read
\begin{equation}
    \label{eq:app_G_prw2}
    \begin{split}
        &G^b(x,y,w) = \lr{(1-r)y + rw}^{z-1},~~~~ G^f(x,y,w) = \frac{1}{S}\lrb{\lr{ax+by+cw}^{z-1} - \lr{by+cw}^{z-1}},\\
        &G^d(x,y,w) = \frac{(1-r)y+rw}{x}\,G^f(x,y,w),
    \end{split}
\end{equation}
where $G^d$ simply accounts for the free child of the $d$-type node that received an obstacle during the previous cascade.
Since $\int_0^1 t^{\nu^s+\nu^f-1}\,dt = (\nu^s+\nu^f)^{-1}$, the pushing weights are obtained from $G^i$ as 
\begin{equation}
    \label{eq:app_pgf_kernel_prw2}
    f^i(\bbs) = G^i(0,0,1) + h(\bbs)\int_0^1 \partial_x G^i(ts_1,ts_2,s_3)\,dt + s_4\int_0^1 \partial_y G^i(ts_1,ts_2,s_3)\,dt,
\end{equation}
where $G^i(0,0,1) = \PP\lrb{\nu^s+\nu^f=0}$ is the childless probability.
For type $b$, $G^b(0,0,1) = r^{z-1}$, and Eq.~\eqref{eq:app_pgf_kernel_prw2} yields, as for the $m$ type of PRW$_1$,
\begin{equation}
    \label{eq:app_pgf_b_prw2}
    f^b(\bbs) = r^{z-1} + \frac{s_4}{s_2}\lrb{\lr{(1-r)s_2 + r s_3}^{z-1} - \lr{r s_3}^{z-1}}.
\end{equation}
For type $f$, $G^f(0,0,1) = 0$ since $\xi^s>0$, and the integrals are explicit,
\begin{equation}
    \label{eq:app_pgf_f_prw2}
    f^f(\bbs) = \frac{1}{S}\lrb{\frac{a h(\bbs) + b s_4}{a s_1 + b s_2}\lr{g(\bbs)^{z-1} - \lr{c s_3}^{z-1}} - \frac{s_4}{s_2}\lr{\lr{b s_2 + c s_3}^{z-1} - \lr{c s_3}^{z-1}}}.
\end{equation}
Finally, for type $d$, $G^d(0,0,1) = p$ is the dead-end probability already met in Eq.~\eqref{eq:movd_prw2}, and
\begin{equation}
    \label{eq:app_pgf_d_prw2}
    f^d(\bbs) = p + h(\bbs)\int_0^1 \partial_x G^d(ts_1,ts_2,s_3)\,dt + s_4\int_0^1 \partial_y G^d(ts_1,ts_2,s_3)\,dt.
\end{equation}
Here the integrals produce finite sums rather than a compact closed form. In practice, the integrands are polynomials of degree at most $z-2$ in $t$, which are numerically evaluated whenever needed.

\section{Small-$\rho$ expansion of the PRW$_2$ escape probability}
\label{app:smallrho_prw2}

We provide the derivation of Eq.~\eqref{eq:smallrho_prw2}. As $\rho\to 0$, we have $a\to 1$, $b\simeq\rho$, $c = \rho^{z}$, $r = \rho^{z-1}$ and $S\to 1$. As for PRW$_1$, we look for extinction probabilities $\bbq = (q^s,q^f,q^b,q^d)$ vanishing as powers of $\rho$, and solve the fixed-point equations $q^i = f^i(\bbq)$ of Appendix~\ref{app:pgf_prw2} to lowest order. Throughout, we use the two auxiliary pgfs introduced there, $g(\bbs) = a s_1 + b s_2 + c s_3$ and $h(\bbs) = (1-r)s_2 + r s_3$, which respectively generate the type of a child with untouched Lorentz statistics and of an occupied node with untouched children. 
We postulate the hierarchy ansatz $q^s, q^f \ll q^d \ll q^b$, which we check a posteriori.
\begin{enumerate}
    \item \textit{Type $b$.} In Eq.~\eqref{eq:app_pgf_b_prw2} evaluated at $\bbs = \bbq$, the bracket is bounded by $(z-1)(1-r)q^f\, h(\bbq)^{z-2}$, so that the second term is at most $(z-1)q^d h(\bbq)^{z-2}$ and is subleading. Hence
    \begin{equation}
        q^b \simeq r^{z-1} = \rho^{(z-1)^2}:
    \end{equation}
    to lowest order, a $b$-type node goes extinct when it is a dead end, that is, when all its children are themselves of type $b$.
    \item \textit{Type $d$.} Similarly, one checks that all configurations other than the dead end require at least one further subtree to go extinct, and are subleading, so that
    \begin{equation}
        q^d \simeq p = \frac{(z-1)ac^{z-2}r}{S} \simeq (z-1)\rho^{z^2-z-1}.
    \end{equation}
    The corresponding event is that the single free child of the $d$-type node received the obstacle and became of type $b$, all its other children being of type $b$ as well.
    \item \textit{Type $s$.} The fixed-point equation reads $q^s = g(\bbq)^{z-1}$, with $g(\bbq) = a\,q^s + b\,q^f + c\,q^b$. Provided $a\,q^s + b\,q^f \ll c\,q^b \simeq \rho^{z^2-z+1}$, we obtain
    \begin{equation}
        q^s \simeq (c\,q^b)^{z-1} = \rho^{(z-1)(z^2-z+1)}.
    \end{equation}
    \item \textit{Type $f$.} Expanding Eq.~\eqref{eq:app_pgf_f_prw2} to first order in $a\,q^s + b\,q^f \ll c\,q^b$, the two contributions proportional to $b\,q^d$ cancel, and, using $h(\bbq) \simeq r q^b$,
    \begin{equation}
        q^f \simeq (z-1)(c\,q^b)^{z-2}\, r\, q^b = (z-1)\rho^{(z-1)(z^2-z+1)-1}.
    \end{equation}
    In words, the $f$-type node has a single free child, which receives the obstacle and becomes a $b$-type dead end, while its $z-2$ other children are $b$-type dead ends.
\end{enumerate}
For $z\geq 3$, both $q^s$ and $q^f$ are indeed much smaller than $\rho^{z^2-z}$, which validates the assumption $a\,q^s + b\,q^f \ll c\,q^b$ used in steps 3 and 4.
Finally, the root pgf \eqref{eq:app_pgf_s_prw2} yields
\begin{equation}
    1 - \PP^{\text{PRW}_2}\lrb{\bbx_n\to\infty} = f^o(\bbq) = g(\bbq)^{z} \simeq (c\,q^b)^{z} = \rho^{z(z^2-z+1)}.
\end{equation}
The exponent has a simple interpretation: to lowest order, PRW$_2$ is trapped when all $z$ children of the root are $b$-type dead ends, which requires $z$ obstructed children, $z(z-1)$ obstructed grandchildren and $z(z-1)^2$ obstructed great-grandchildren, that is, $z\lr{1 + (z-1) + (z-1)^2} = z(z^2-z+1)$ obstacles. The same counting with one fewer layer recovers the exponent $z\lr{1+(z-1)} = z^2$ of PRW$_1$, Eq.~\eqref{eq:smallrho_prw1}.
Following this pysical picture, the extrapolation to higher $k$ orders is immediate:
\begin{equation}
    1 - \PP^{\text{PRW}_k}\lrb{\bbx_n\to\infty} \underset{\rho \to 0}{\sim} \rho^{z\lrb{\sum_{\ell=0}^{k}(z-1)^\ell}}.
\end{equation}

\section{Numerical evaluation of escape probabilities and trace statistics}
\label{app:numerics}

\subsection{Fixed-point solvers}
\label{app:fixedpoint}

All the quantities of Sections~\ref{sec:prw1}, \ref{sec:prw2} and \ref{sec:traces} are fixed points of maps built from the pgfs $\bbf$, which we compute by plain iteration.

\paragraph{Extinction probabilities.} Starting from $\bbq_0 = \boldsymbol{0}$, the iterates
\begin{equation}
    \label{eq:app_fp_ext}
    \bbq_{k+1} = \bbf(\bbq_k)
\end{equation}
increase monotonically to the smallest fixed point $\bbq$ of Eq.~\eqref{eq:fp_prw1}, and $q^i_k = \PP^i\lrb{|\bbZ_k| = 0}$, see Eq.~\eqref{eq:ext_gen}. We stop when $\|\bbq_{k+1}-\bbq_k\|_\infty < \varepsilon$, and the escape probability is then $1 - f^o(\bbq_k)$.
The convergence is geometric, with rate $\varrho\lr{J(\bbq)}<1$ for $\rho\neq\rho_c$, where $J(\bbs)_{ij} = \partial_{s_j}f^i(\bbs)$ is the Jacobian of $\bbf$, with $J(\ind{}) = M^T$. At $\rho = \rho_c$, it becomes algebraic, $\bbq-\bbq_k = O(1/k)$.
The same iterates give the distribution of the maximum depth, Eq.~\eqref{eq:maxdepth_root_gen}, at no extra cost.

\paragraph{Volume and dead-end pgfs.} For fixed $s$ in the closed unit disk of the complex plane, we iterate the functional equations \eqref{eq:rde_vol_pgf} and \eqref{eq:rde_de_pgf},
\begin{equation}
    \label{eq:app_fp_pgf}
    \begin{split}
        &\boldsymbol{\mathcal{N}}_{k+1}(s) = s\,\bbf\lr{\boldsymbol{\mathcal{N}}_k(s)} + (1-s)\bbp_0,~~~~\boldsymbol{\mathcal{N}}_0(s) = \boldsymbol{0},\\
        &\boldsymbol{\mathcal{L}}_{k+1}(s) = \bbf\lr{\boldsymbol{\mathcal{L}}_k(s)} + (s-1)\boldsymbol{\ell},~~~~\boldsymbol{\mathcal{L}}_1(s) = \bbp_0,
    \end{split}
\end{equation}
and the root pgfs follow as $\mathcal{N}^o(s) = s\,f^o\lr{\boldsymbol{\mathcal{N}}_k(s)}$ and $\mathcal{L}^o(s) = f^o\lr{\boldsymbol{\mathcal{L}}_k(s)} + (s-1)\ell^o$.
Note that the dead-end iteration is initialized at $\bbp_0$ rather than $\boldsymbol{0}$.

\subsection{Inversion of pgfs by discrete Fourier transform}
\label{app:fft}

Let $G(w) = \sum_{k\geq 0}p_kw^k$ be any of the pgfs above, e.g.\ $\mathcal{N}^o$ or $\mathcal{L}^o$, which is analytic in the open unit disk and continuous on its closure. Taking the circle $|w| = R\leq 1$ as the contour in the residue formula \eqref{eq:pgf_inversion}, and setting $w = Re^{2i\pi\theta}$, yields
\begin{equation}
    \label{eq:app_cauchy}
    p_k = R^{-k}\int_0^1 G\lr{Re^{2i\pi\theta}}e^{-2i\pi k\theta}\,\der\theta.
\end{equation}
Discretizing this integral with the trapezoidal rule \cite{abate1992numerical} on $J$ equispaced nodes $w_j = R\,\omega^j$, with $\omega = e^{2i\pi/J}$, yields the approximation
\begin{equation}
    \label{eq:app_dft}
    \hat{p}_k = \frac{R^{-k}}{J}\sum_{j=0}^{J-1}G(w_j)\,\omega^{-jk},~~~~0\leq k<J,
\end{equation}
which is nearly the discrete Fourier transform of the $J$ values $G(w_j)$, so that all $\hat{p}_k$ are obtained at once by a fast Fourier transform, in $O(J\log J)$ operations. Each value $G(w_j)$ requires one fixed-point solve \eqref{eq:app_fp_pgf}.
The normalization $\sum_k\hat{p}_k\simeq 1$ and the mean $\sum_k k\,\hat{p}_k\simeq G'(1)$ provide simple consistency checks. Close to $\rho_c$, the distributions develop heavy tails and $J$ must be increased accordingly.

\end{document}